\documentclass[prb,twocolumn,showpacs,amsmath,amssymb,floatfix]{revtex4-1}
\usepackage{graphicx}
\usepackage{color}
\usepackage{pifont}
\usepackage{amstext}
\usepackage{amsmath}
\usepackage{graphics,amssymb}
\usepackage{bbm}
\usepackage{tabularx}

\DeclareMathOperator{\ii}{i}

\begin{document}
\title{Fermi surface topology of the two-dimensional Kondo lattice model: a dynamical cluster approximation approach}
\author{L.~C.~Martin, M.~Bercx and F.~F.~Assaad}
\affiliation{ Institut f\"ur Theoretische Physik und Astrophysik,
Universit\"at W\"urzburg, Am Hubland, D-97074 W\"urzburg, Germany }
\begin{abstract}
We report the results of extensive dynamical cluster approximation calculations, based on a quantum Monte Carlo solver, for the two-dimensional Kondo lattice model. Our particular cluster implementation renders possible the simulation of spontaneous antiferromagnetic 
symmetry breaking.  By explicitly computing the single-particle spectral function  both in the paramagnetic and antiferromagnetic phases,  we follow the evolution of the Fermi surface across this  magnetic transition. The results, computed for clusters up to $16$ orbitals, show clear evidence for the existence of three distinct Fermi surface topologies.   
The transition from the paramagnetic metallic phase to the antiferromagnetic metal is  continuous;  Kondo screening does not break down  and  we observe  a back-folding of the paramagnetic heavy fermion band.  Within the antiferromagnetic phase and when the ordered moment becomes {\it large}  the  Fermi surface evolves to one which is adiabatically connected to a Fermi surface where the local moments are  frozen in an antiferromagnetic order.
\end{abstract}
\maketitle
\section{Introduction}
\label{sec:section1}
Heavy-fermion systems \cite{Stewart_rev, Lee86} are characterized by a hierarchy of distinctive energy scales \cite{Yang08, Burdin10,Zhu10}.
The Kondo scale, $ T_K \propto e ^{-W/J}$ with bandwidth $W$ and superexchange $J$, marks the screening of local magnetic moments. This screening is a many-body effect which entangles 
the spins of the conduction electrons and local moments \cite{Kondo64Yosida66}.
Below the coherence temperature, which is believed to track the Kondo scale \cite{Georges00, Assaad04a},  the paramagnetic (PM)
heavy-fermion liquid \cite{Lohneysen_rev} emerges and corresponds to a coherent, Bloch like, superposition
of the screening clouds of the individual magnetic moments. Even in the Kondo limit, where
charge fluctuations of the impurity spins are completely suppressed, this paramagnetic state is
characterized by a large Fermi surface with Luttinger volume including both the magnetic moments
and conduction electrons \cite{Martin82, Oshikawa00, Lee08}. The coherence temperature of this metallic state is small
or, equivalently, the effective mass large.\\
Kondo screening competes with the Ruderman-Kittel-Kasuya-Yosida (RKKY) interaction,
which indirectly couples the local moments via the magnetic polarization of the conduction
electrons. The RKKY energy scale is set by $J^{2} \chi_{c}({\mathbf q},\omega = 0) $ where  $\chi_{c} $ corresponds to the spin susceptibility of the conduction electrons \cite{Hewson_book}.\\
The competition between Kondo screening - favoring paramagnetic ground states - and the RKKY interaction - favoring magnetically ordered states - is at the heart of quantum
phase transitions \cite{Doniach77, Varma76}, the detailed understanding of which is still under debate (for
recent reviews see Ref. \onlinecite{Gegenwart08, Si10, Yamamoto10, Vojta10}).\\
Here, two radically different scenarios have been put forward to
describe this quantum phase transition. In the \textit{standard} Hertz-Millis picture \cite{Hertz76, Millis93}, the quasi-particles
of the heavy-fermion liquid remain intact across the transition and undergo a spin-density wave transition.   In particular, neutron scattering experiments of  the heavy-fermion system 
$\mbox{Ce}_{1-x}\mbox{La}_{x}\mbox{Ru}_{2}\mbox{Si}_{2}$ show that fluctuations of the antiferromagnetic order parameter are responsible for the magnetic phase transition and that the transition is well understood in terms of the Hertz-Millis approach \cite{Knafo09}.\\
On the other hand, since many experimental observations such as the almost wave vector
independent spin susceptibility in $\mbox{CeCu}_{6-x}\mbox{Au}_{x}$ \cite{Schroeder00}, or the jump in the low-temperature Hall
coefficient in $\mbox{YbRh}_{2}\mbox{Si}_{2}$ \cite{Paschen04} are not accounted for by this theory alternative scenarios have
been put forward \cite{Si01, Senthil04, Watanabe07, Watanabe09, Lanata08}. 
In those scenarios,  the quantum critical point is linked to the very breakdown of the  quasi-particle of the
heavy-fermion state \cite{Coleman01, Klein09}, and a topological reorganization of the Fermi surface across the transition
is expected \cite{Paschen04, Klein08}. \\
Recent experiments on $\mbox{CeIn}_{3}$ \cite{Harrison07} or $\mbox{CeRh}_{1-x}\mbox{Co}_{x}\mbox{In}_{5}$ \cite{Goh08} show that a change in Fermi surface (FS) topology must not necessarily occur only at the magnetic order-disorder quantum critical point (QCP). In fact, even  in YbRh$_{2}$Si$_{2}$ it has since been shown that the Fermi surface reconstruction can be shifted to either side of the QCP via application of positive or negative chemical pressure \cite{Friedemann09}.

In this paper, we address the above questions through an explicit calculation of the Fermi surface topology in the framework of the Kondo lattice model (KLM). In its simplest form the KLM describes an array of localized magnetic moments of spin $1/2$, arising from atomic $f$-orbitals, that are coupled antiferromagnetically (AF) via the exchange interaction $J$ to a metallic host of mobile conduction electrons.\\
We present detailed dynamical cluster approximation (DCA) calculations aimed at the investigation of the KLM ground state. For the simulations within the magnetically ordered phase, we have extended the DCA to allow for symmetry breaking antiferromagnetic order. We map out the magnetic phase diagram as a function of $J/t$ and conduction electron density $n_{c}$, with particular interest in the single-particle spectral function and the evolution of the Fermi surface. 
The outline is as follows. The model and the DCA implementation is discussed in Sec.~\ref{sec:section2}. Results for the case of half-band filling and hole-doping are discussed in Sec.~\ref{sec:section3} and \ref{sec:section4}. Section  ~\ref{sec:section5} is devoted to a summary. 
This paper is an extension to our previous work, where part of the results have already been published \cite{Lee08}.
\section{Model Hamiltonian and dynamical cluster approximation}
\label{sec:section2}
The Kondo lattice model (KLM) we consider reads
\begin{equation}
H=  \sum_{ {\mathbf k},\sigma } (\epsilon( {\mathbf k}  ) -\mu) c^{\dagger}_{ {\mathbf k}, \sigma} c_{ {\mathbf k}, \sigma } +
 J \sum_{\mathbf i }  {\mathbf S}^{c}_{\mathbf i} \cdot {\mathbf S}^{f}_{\mathbf i}\;.
\label{eqn:eqn1}
\end{equation}
The operator $c^{\dagger}_{ {\mathbf k},\sigma}$ denotes creation of an electron in a Bloch state with wave vector ${\mathbf k}$ and a z-component of spin $\sigma=\uparrow , \downarrow$. The spin $1/2$ degrees of freedom, coupled via $J>0$, are represented with the aid of the Pauli spin matrices ${\pmb \sigma}$ by 
${\mathbf S}^{c}_{\mathbf i}=\frac{1}{2} \sum_{s,s'} c^{\dagger}_{{\mathbf i}, s} {\pmb \sigma}_{s,s'} c_{ {\mathbf i}, s'}$ and the equivalent definition for ${\mathbf S}^{f}_{ \mathbf i}$ using the localized orbital creation operators $f^{\dagger}_{{\mathbf i},\sigma}$. The chemical potential is denoted by $\mu$. The definition of the KLM  excludes charge fluctuations on the $f$-orbitals and as such a strict constraint of one electron per localized $f$-orbital has to be included.  For an extensive review of this model we refer the reader to Ref. \onlinecite{Tsunetsugu97_rev}. \\
Particle-hole symmetry at half-filling is given if hopping is restricted to nearest neighbors on the square lattice and the chemical potential is set to zero. We introduce a next-nearest neighbor hopping with matrix element $t'$ to give a modified dispersion $\epsilon ({\mathbf k})= -2 t \left[ \cos(k_{x}) + \cos(k_{y})\right] -2 t' \left[ \cos(k_{x}+k_{y}) + \cos(k_{x}-k_{y})\right]$. As we will see,  the suppression  of particle-hole symmetry due to a finite value of $t'$ leads to  dramatic changes in the   scaling of the quasi-particle gap at low values of $J/t$ and at half-band filling. We have considered the value $t'/t= -0.3$. This choice guarantees that the spin susceptibility of the host metallic state at small dopings away from half-filling   is peaked at wave vector ${\mathbf Q }= (\pi,\pi)$. Thus, antiferromagnetic  order  as opposed to an incommensurate spin state is favored.\\

\begin{figure}
\begin{center}
\includegraphics[width=\columnwidth,type=png,ext=.png,read=.png]{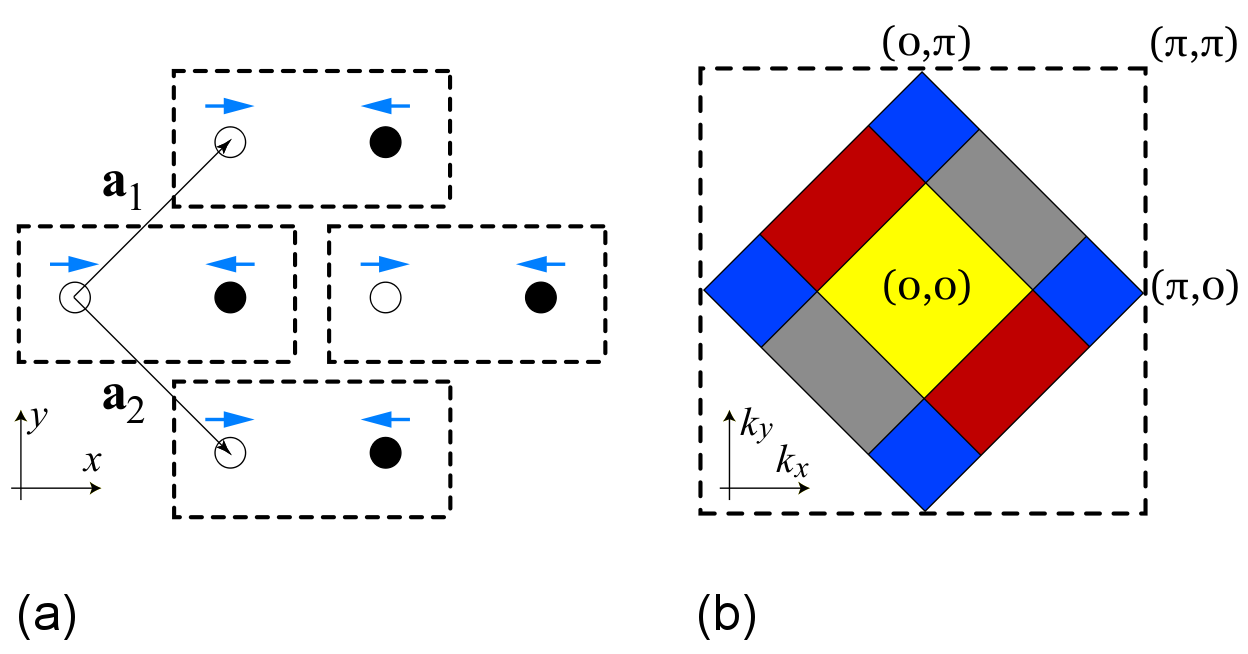}
\end{center}
\caption{(Color online) Definition of the antiferromagnetic unit cell (a)  and DCA patching of the Brillouin zone (b). The lattice vectors  $\mathbf{a}_{1}=(1,1)$ and $\mathbf{a}_{2}=(1,-1)$ connect the AF unit cells. The two inequivalent $c$-orbitals are denoted by filled (empty) circles and the localized $f$-orbitals with arrows. The reciprocal-space lattice vectors $\mathbf{b}_{1}$ and $\mathbf{b}_{2}$ span the magnetic Brillouin zone. The MBZ is uniformly discretized in $N_{p}=4$ patches, corresponding to real-space clusters each containing $N_{p}$ magnetic unit cells. The color coding refers to constant momentum dependency of the self-energy and cluster Green function. The dashed square denotes the extended Brillouin zone.}
\label{fig:fig1}
\end{figure}

To solve the model we have used the DCA \cite{Hettler00,Maier05} approach which retains \textit{ temporal } fluctuations and hence accounts for the Kondo effect   but neglects spatial fluctuations on a length scale larger than the cluster size. The approach relies on the coarse-graining of momentum space, and momentum conservation holds only between the ${\mathbf k}$-space patches.  By gradually defining smaller sized patches the DCA allows to restore the ${\mathbf k}$-dependency of the self-energy. \\
The standard formulation of DCA is naturally translationally invariant. To measure orders beyond translation invariance the DCA can be generalized to lattices with a supercell containing  $N_{u}$-unit cells of the original lattice \cite{Maier05}.
A general unit cell is addressed by ${\bar{\mathbf x}} ={\mathbf x} +{\mathbf r}_{\mu} $, ${\mathbf x}$ denoting the supercell and ${\mathbf r}_{\mu}$ the relative points with $\mu=1...N_{u}$.
In this work we opted for  $N_{u}=2$ which is the minimum requirement for the formation of antiferromagnetic order. The reduced Brillouin zone (RBZ) is then spanned by $\mathbf{b}_{1}=\pi(1,1)$ and $\mathbf{b}_{2}=\pi(1,-1)$ and we have set the lattice constant to unity. 
This  supercell as well as the RBZ  (also coined magnetic Brillouin zone (MBZ)) are plotted in  Fig. \ref{fig:fig1}. \\
The self-energy and Green function are to be understood as spin dependent matrix functions, with an index for the  unit cell within the supercell as well as an orbital index for the $c$- and $f$-orbitals in each unit cell.  The ${\mathbf k}$-space discretization into $N_{p}$ patches with momentum conservation only between patches yields the coarse-grained lattice Green function 
\begin{eqnarray}
\bar{G}_{\sigma}({\mathbf K},\ii \omega) = \frac{N_{p}}{N} \sum_{\tilde{{\mathbf k}}} G_{\sigma}({\mathbf K}+\tilde{{\mathbf k}},\ii \omega)
\label{eqn:eqn2}
\end{eqnarray}
and ensures that the  self-energy is only dependent on  the coarse-grained momentum ${\mathbf K} $:  $ \bar{\Sigma} \equiv \bar{\Sigma}_{\sigma}({\mathbf K},\ii \omega)$. Here, the reciprocal vector ${\mathbf K}$ denotes the center of a patch and the original ${\mathbf k}$ vectors are given by ${\mathbf k}={\mathbf K}+\tilde{{\mathbf k}}$.
The self-energy is extracted  from a real-space cluster calculation  with periodic boundaries  yielding the  quantized ${\mathbf K}$ values.  Let $ {\cal G}_{0,\sigma}({\mathbf K},\ii \omega) $ be  the bath (non-interacting) Green function  of  the  cluster  problem  and  ${\cal G}_{\sigma}({\mathbf K},\ii \omega)$ the full cluster  Green function.  Hence, $\Sigma={\cal G}_{0}^{-1}-{\cal G}^{-1}$ and self-consistency requires that:
\begin{eqnarray}
{\cal G}_{\sigma}  ({\mathbf K},\ii \omega) 
& = & ({\cal G}^{-1}_{0,\sigma}({\mathbf K},\ii \omega) -\bar{\Sigma}_{\sigma}({\mathbf K},\ii \omega) )^{-1}\\
& = & {\frac{N_{p}}{N}}
\sum_{\tilde{{\mathbf k}}}
( G_{0,\sigma}^{-1}
 ({\mathbf K}+\tilde{{\mathbf k}},\ii \omega) - \bar{\Sigma}_{\sigma}({\mathbf K},
 \ii \omega))^{-1}.  \nonumber
\label{eqn:eqn3}
\end{eqnarray}\\
The non-interacting lattice Green function is denoted by 
$ G_{0,\sigma}({\mathbf K}+\tilde{{\mathbf k}},\ii \omega)$.\\
To summarize, our implementation of DCA approximates the Fourier space of a lattice with two-point basis by patching. Regarding the KLM this leads to real-space clusters of size $N_{p}$, each encompassing $2\times N_{p}$ $c$- and $f$-orbitals. In the present work cluster sizes $N_{p}=1$ and  $N_{p}=4$  are considered (Fig. \ref{fig:fig1}).\\
The lattice green function $G_{\sigma}({\mathbf k}, \ii \omega_{n})$ with ${\mathbf k} \in  MBZ $ can equally be expressed as $g_{\sigma}(\bar{{\mathbf k}},\ii \omega_{n})$ in an extended Brillouin zone scheme, with $\bar{\mathbf k} \in BZ $ and $\bar{\mathbf k}={\mathbf k} +m\mathbf{b}_{1}+n\mathbf{b}_{2}$:
\begin{eqnarray}
g_{\sigma}(\bar{{\mathbf k}},\ii \omega_{n})=\frac{1}{2}\sum_{\mu, \nu}^{N_{u}=2} e^{ \ii \bar{{\mathbf k}}({\mathbf r}_{\nu}- {\mathbf r}_{\mu})} \left[ G_{\sigma}({\mathbf k}, \ii \omega_{n})\right]_{\mu \nu}\;.
\label{eqn:eqn4}
\end{eqnarray}
In order to implement the DCA one has to be able to solve, for a given bath, the Kondo model on the effective cluster. The method we have chosen is the auxiliary field Quantum Monte Carlo (QMC) version of the Hirsch-Fye algorithm, following precisely the same realization as in Ref. \onlinecite{Capponi00}.    The implementation details of the self-consistency cycle   can be found in Ref. \onlinecite{Beach08}.
The performance of the QMC cluster solver has been enhanced by almost an order of magnitude by implementing the method of delayed updates \cite{Alvarez08}. During the QMC Markov process the full updates of the Green function are delayed until a sufficiently large number of local changes in the auxiliary field have been accumulated. This enhances the performance of the whole QMC algorithm since considerable CPU time
is spent in the updating section. To efficiently extract spectral information from the imaginary time discrete QMC data a stochastic analytic continuation scheme is employed \cite{Beach04a}.\\
Ground state properties of the model are accessed by extrapolation of the  inverse temperatures $\beta$  to infinity. This is a demanding task since all relevant energy scales, the Kondo scale, the coherence scale and the RKKY scale, become dramatically smaller with decreasing $J/t$. In the paramagnetic phase and below the coherence scale a clear hybridization gap is apparent in the single particle spectral function and we use this criterion to estimate the coherence temperature. 
Since the computational time required by the QMC cluster solver \cite{HirschFye86} increases proportional to $(\beta N_{u}N_{p})^3$, this limits us in the resolution of energy scales to values of $J/t \geq 0.8$.   It is important to note that for small dopings away from half-band filling and for the considered cluster sizes, the negative sign problem is not  severe and hence is not the limiting factor.
\section{The Half-Filled KLM}
\label{sec:section3}
\begin{figure}
\begin{center}
\includegraphics[width=\columnwidth,type=png,ext=.png,read=.png]{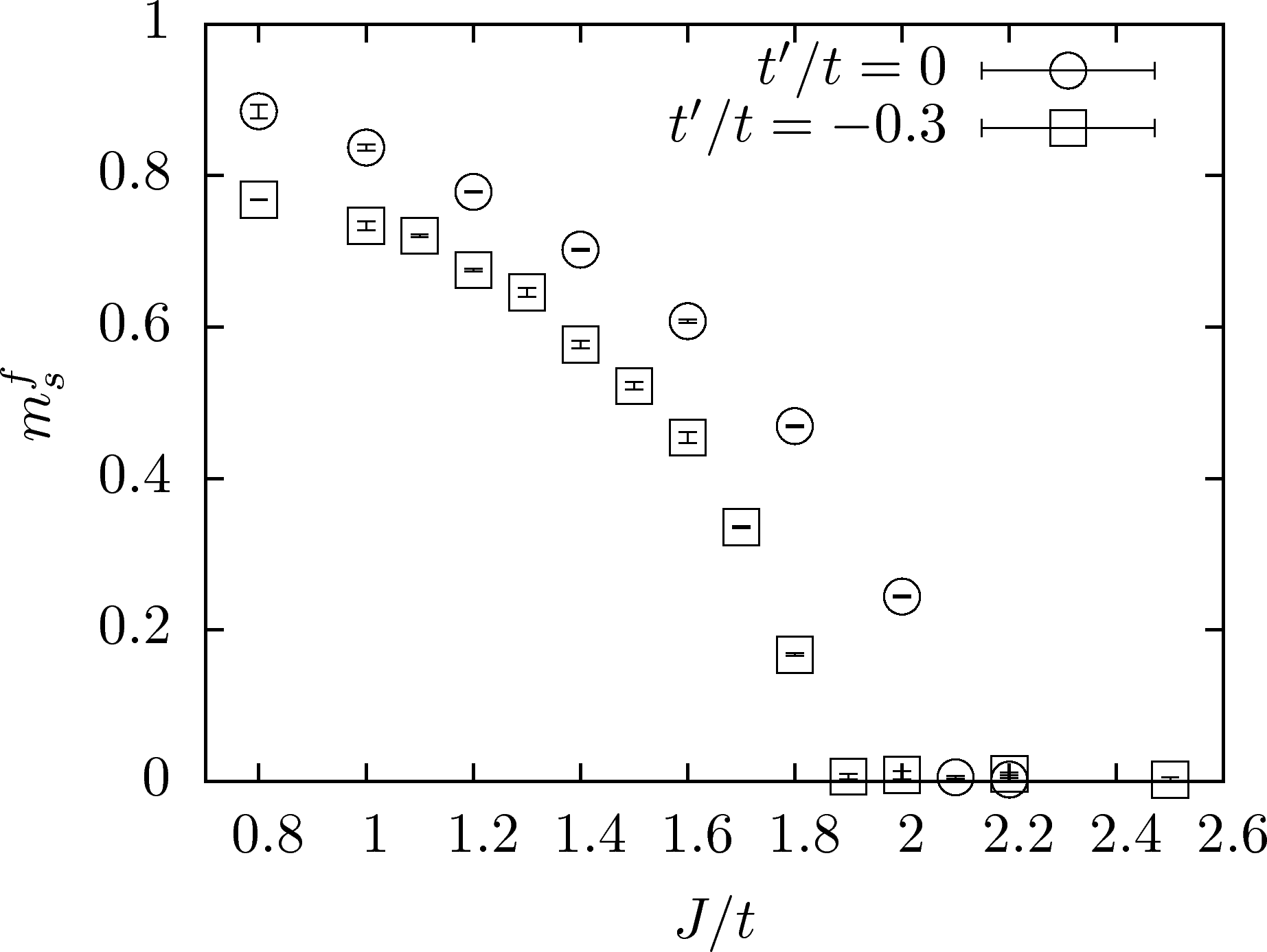} 
\end{center}
\caption{The staggered magnetization $m^{f}_{s}$ of the local moment spins in the ground state as a function of coupling $J/t$ with next-nearest neighbor hopping $t'/t=-0.3$ and $t'/t=0$, respectively. The low temperature limit is reached by performing simulations at various inverse temperatures $\beta$ . Convergence to the ground state has been achieved for the following inverse temperatures:
 $\beta t=20$ ($2.5 \geq J/t \geq 1.7$), $\beta t=40$  ($1.6 \geq J/t \geq  1.1$), $\beta t=80$ ($J/t=1$) and $\beta t=100$ ($J/t=0.8$).}
\label{fig:fig2}
\end{figure}

We consider the KLM at half-filling in two cases: either with or without particle-hole symmetry. That is  
$t'/t = 0$ and $t'/t = -0.3$ respectively. 
At $t'=0$ lattice QMC  simulations do not suffer  from the negative sign problem, and it is well established that   the 
Kondo screened phase gives way to an AF ordered phase at $J_c/t = 1.45$ \cite{Capponi00,Assaad99a}.  This magnetic order-disorder 
 transition  occurs  between insulating states and it is reasonable to assume that it belongs to the three-dimensional  
$O(3)$  universality class \cite{Troyer97}.   Here, the dynamical exponent takes the value of unity such that  the correlation lengths in imaginary time and in real space are locked in together and diverge at the critical point.  Due to the very small cluster sizes  considered in the DCA it is clear that we will not capture the physics of this transition. In fact as soon as the correlation length exceeds the  size of the DCA cluster,  symmetry breaking signaled by a finite value of the staggered moment
\begin{equation}
 m^{f}_{s}=\frac{1}{2N_{p}} \sum_{\mathbf i} \langle n^{f}_{{\mathbf i},\uparrow} - n^{f}_{{\mathbf i},\downarrow} \rangle e^{- \ii {\mathbf Q} {\mathbf i}}
\label{eqn:eqn5}
\end{equation}
sets in and mean-field exponents are expected. The normalization is chosen such that the  staggered magnetization of the fully polarized state takes the value of unity.
The above quantity is plotted in Fig.~\ref{fig:fig2}  at $t'/t = 0$ and  $N_{p}=1$   As apparent,  
the critical value of $J/t$ overestimates ( $J_{c}/t \approx 2.1$.)  the {\it exact   } lattice QMC result. Switching off nesting by  including a  finite value of $t'/t$  shifts the magnetic order-disorder transition to lower values of $J/t$.\\
To improve on this result, we can systematically enhance the cluster size. However, our major interest lies in the  single-particle 
spectral function. As we will see below, and well within the magnetically ordered phase, this quantity compares very well with the {\it exact} lattice QMC results.
\subsection{Single-particle spectrum - $t'/t=0$}
\label{sec:subsection3A}
\begin{figure}
\begin{center}
\includegraphics[width=\columnwidth,type=png,ext=.png,read=.png]{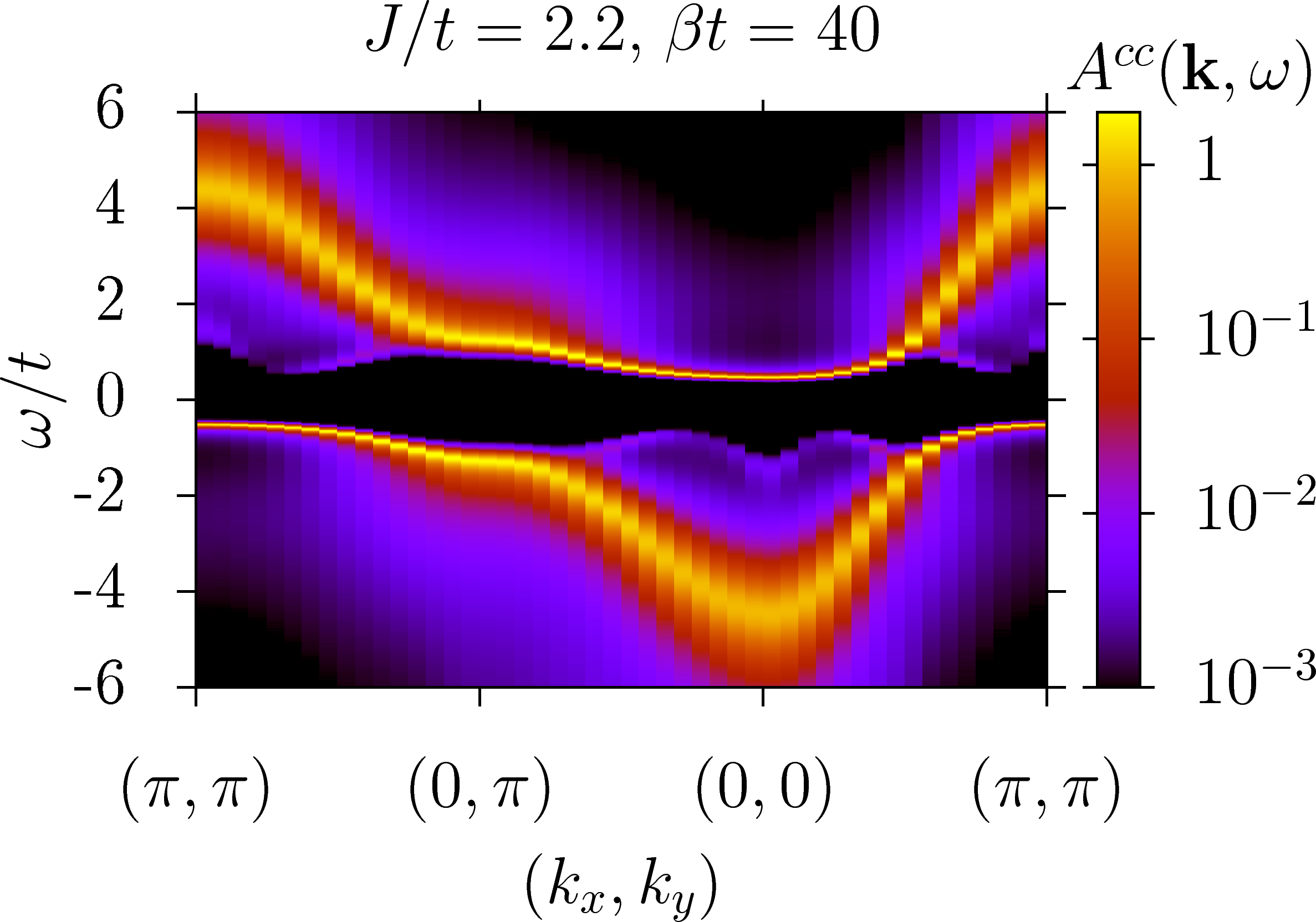}
\end{center}
\caption{(Color online) Single-particle spectral function $A^{cc}({\mathbf k}, \omega)$ at half-filling ($n_{c} = 1$) and with particle-hole symmetry ($t'/t=0$) close to the magnetic phase transition on the paramagnetic side, $J/t=2.2$.}
\label{fig:fig3}
\end{figure}

\begin{figure}
\begin{center}
\includegraphics[width=\columnwidth,type=png,ext=.png,read=.png]{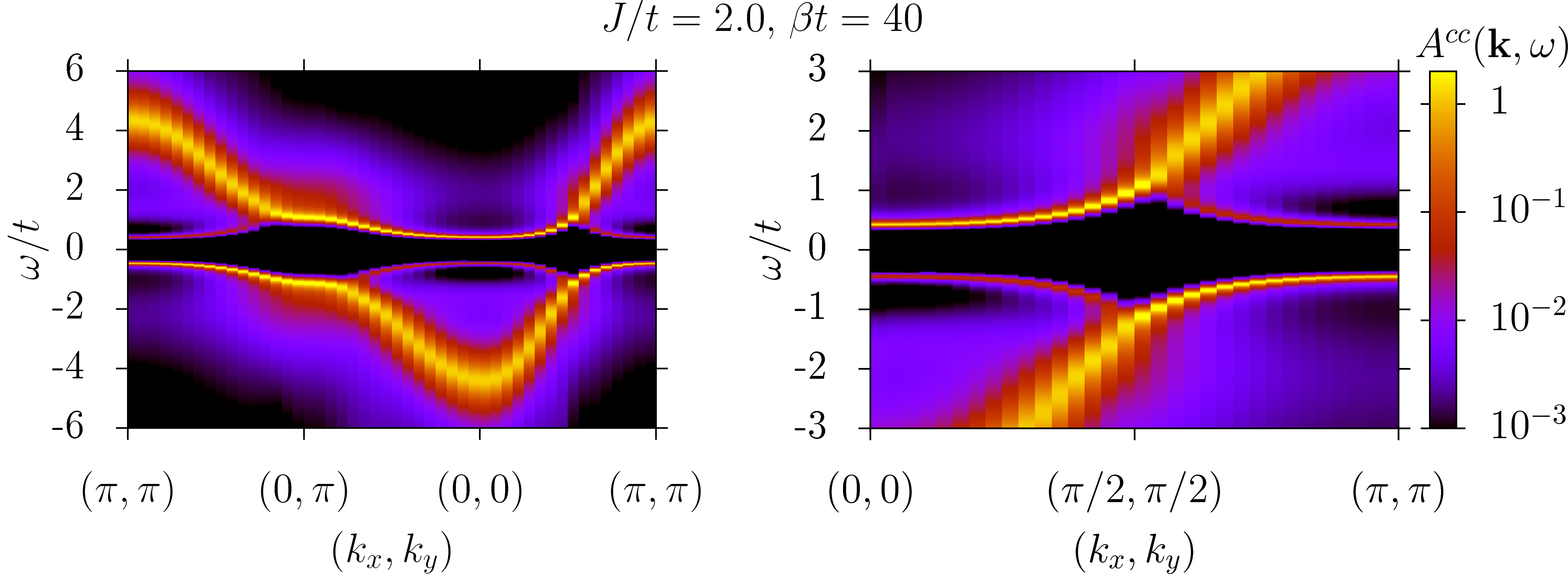}
\caption{(Color online) Single-particle spectral function $A^{cc}({\mathbf k}, \omega)$ at half-filling ($n_{c} = 1$) and with particle-hole symmetry ($t'/t=0$) close to the magnetic phase transition on the AF side, $J/t=2.0$ (a) and closeup (b).  The onset of magnetic order and concomitant emergence of the MBZ is signaled by the appearance of shadow bands.  The staggered magnetization for this point is measured to be $m^{f}_{s}=0.244 \pm 0.001$.}
\label{fig:fig4}
\end{center}
\end{figure}

\begin{figure}
\begin{center}
\includegraphics[width=\columnwidth,type=png,ext=.png,read=.png]{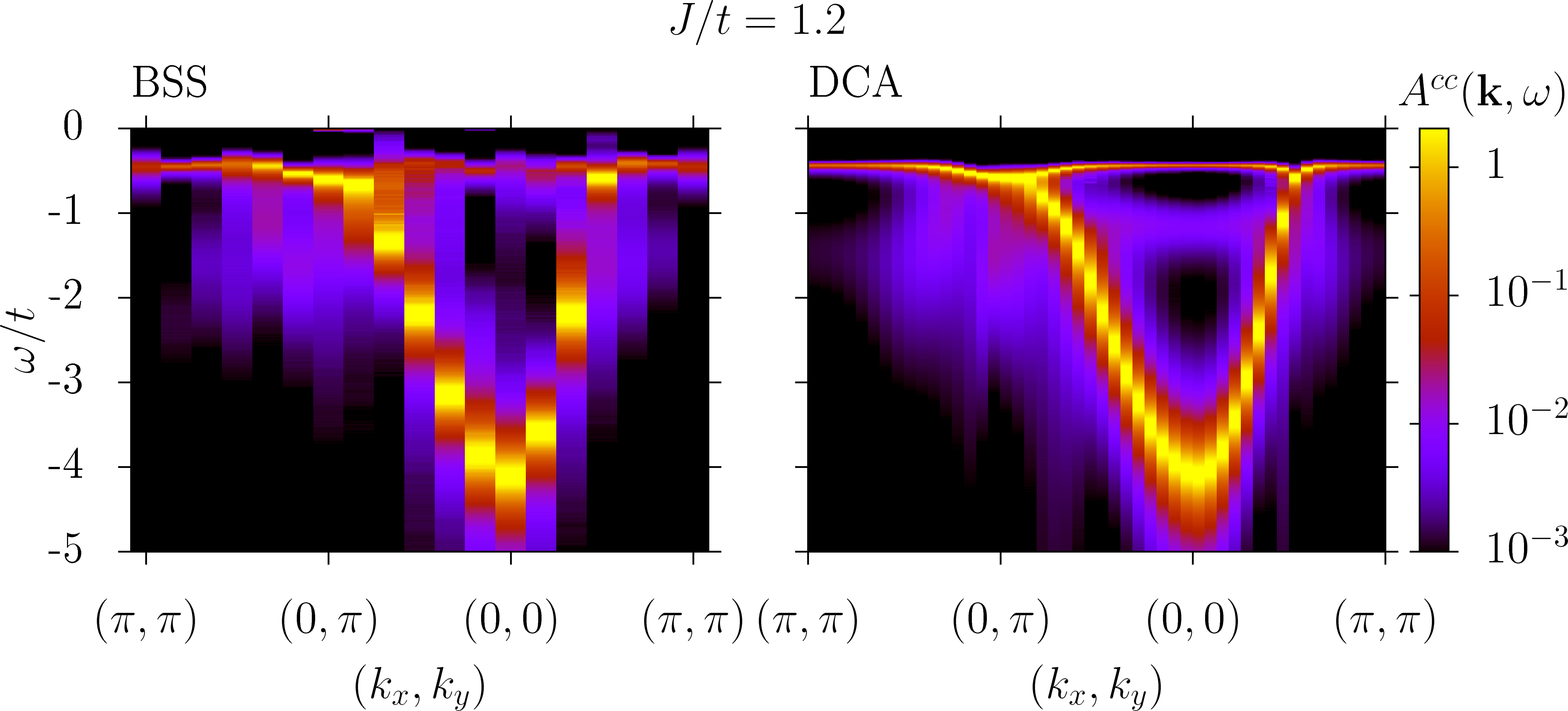}  \\
\end{center}
\caption{(Color online) Comparison of the DCA single-particle spectral function $A^{cc}({\mathbf k}, \omega)$  with the $T=0$ spectral function. 
Data (a) taken from Ref. \onlinecite{Capponi00} which used a projective auxiliary field Monte Carlo (BSS) method for the KLM on a $12 \times 12$ lattice and our DCA results (b) at $\beta t=40$, both for the same parameter set - $t'/t=0$, $n_{c}= 1$, $J/t=1.2$.}
\label{fig:fig5}
\end{figure}

We have calculated the single-particle spectral function 
$A^{cc}({\mathbf k}, \omega)= -\frac{1}{\pi}  \sum_{\sigma} \mbox{Im} \left [g^{cc}_{\sigma} ( {\mathbf k}, \omega)\right ] $ of the conduction electrons along a path of high symmetry in the extended Brillouin zone.
In all our spectral plots, on the energy axis, $\omega $ values are given relative to the chemical potential $\mu$.\\
In Fig.~\ref{fig:fig3} we plot the spectrum for $J/t=2.2$  which was seen to be in the paramagnetic phase (see Fig.~\ref{fig:fig2}). The lower band runs very flat around ${\mathbf k}=(\pi,\pi)$, with relatively low spectral weight (note the logarithmic scale of the color chart) in comparison to the other parts of the band which are mostly unchanged from the non-interacting case. This feature is associated with Kondo screening of the impurity spins and the resultant large effective mass of the composite quasi-particles. Since no band crosses the Fermi energy ($\omega/t=0$) we classify this region of parameter space as a Kondo insulator. The observed dispersion relation  is well described already in the framework of the large-$N$ mean-field  theory  of the Kondo lattice model 
\cite{Georges00}.  At the particle-hole symmetric point this approximation, which in contrast to the  DCA+QMC results of Fig.~\ref{fig:fig3}  neglects the constraint of no double   occupancy of the $f$-orbitals, yields the dispersion relation of hybridized bands:
\begin{equation}
E_{\pm}(\mathbf{k}) = \frac{1}{2}  \left[  \epsilon(\mathbf{k}) \pm  \sqrt{ \epsilon(\mathbf{k})^2 + \Delta^2 }\right].
\label{eqn:eqn6}
\end{equation}
Within the mean-field theory the  quasi-particle gap (see  Eq. \ref{eqn:eqn7}), $\Delta_{qp} \propto \Delta^{2}$, tracks the Kondo scale.\\
At $J/t=2.0$ we have measured a non-vanishing staggered magnetization $m^{f}_{s}=0.244 \pm 0.001$. The spectral function (Fig.~\ref{fig:fig4}) for this lightly AF ordered simulation now includes additional low-energy band structures: In the upper band around ${\mathbf k}=(\pi,\pi)$ and in the lower band around ${\mathbf k}=(0,0)$.  These  {\it shadow bands} arise due to the  scattering of the heavy quasi-particle  off the magnetic fluctuations  centered at  wave vector $\mathbf{Q} = (\pi,\pi)$.\\
At $J/t=1.2$ we make a comparison with the spectrum obtained via QMC lattice simulations using the projective auxiliary field  algorithm of Ref. \onlinecite{Capponi00} for the same parameters. We see in Fig.~\ref{fig:fig5} the QMC lattice simulation results \cite{Capponi00} on the left and our DCA result on the right, both only for the photoemission spectrum $(\omega/t < 0)$. 
The agreement of the DCA spectrum with the Blankenbecler-Scalapino-Sugar (BSS) result is excellent. 
As in the BSS lattice calculations and  for the particle-hole symmetric case, the conduction (valence)  band maximum 
(minimum)  is located at   $ \mathbf{k}= ( \pi,\pi)  $ ( $ \mathbf{k} = ( 0,0) $)  for all  considered values of $J/t$.\\
It is noteworthy that already with the smallest possible cluster capable of capturing AF order, $N_{p}=1$, we are able to produce a single-particle spectrum which is essentially the same as the lattice QMC result. This confirms that the DCA is indeed a well suited approximation for use with the KLM: the essence of the competition between RKKY-mediated spacial magnetic order and the time-displaced correlations responsible for Kondo screening is successfully distilled to a small cluster dynamically embedded in the mean-field of the remaining bath electrons.
\subsection{Single-particle spectrum - $t'/t=-0.3$}
\label{sec:subsection3B}
\begin{figure*}
\begin{center}
\includegraphics[width=\textwidth,type=png,ext=.png,read=.png]{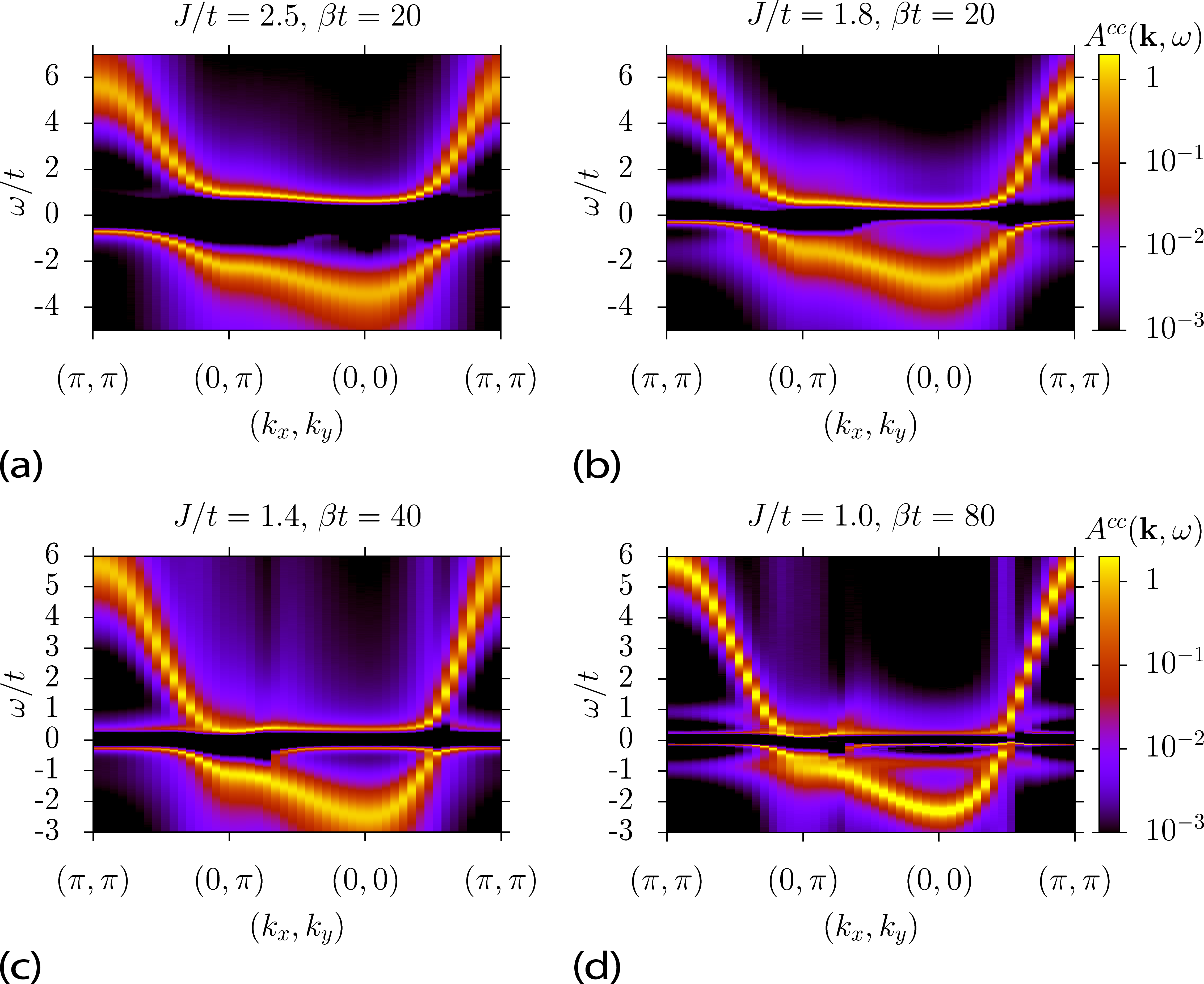}
\caption{(Color online) Single-particle spectral functions $A^{cc}({\mathbf k}, \omega)$ at half-filling ($n_{c} = 1$) with $t'/t=-0.3$. As the Kondo coupling $J/t$ is decreased, the spectrum changes from a paramagnetic Kondo insulator with indirect gap (a) to an antiferromagnetic ordered state (d).}
\label{fig:fig6}
\end{center}
\end{figure*}
\begin{figure*}
\begin{center}
\includegraphics[width=\textwidth,type=png,ext=.png,read=.png]{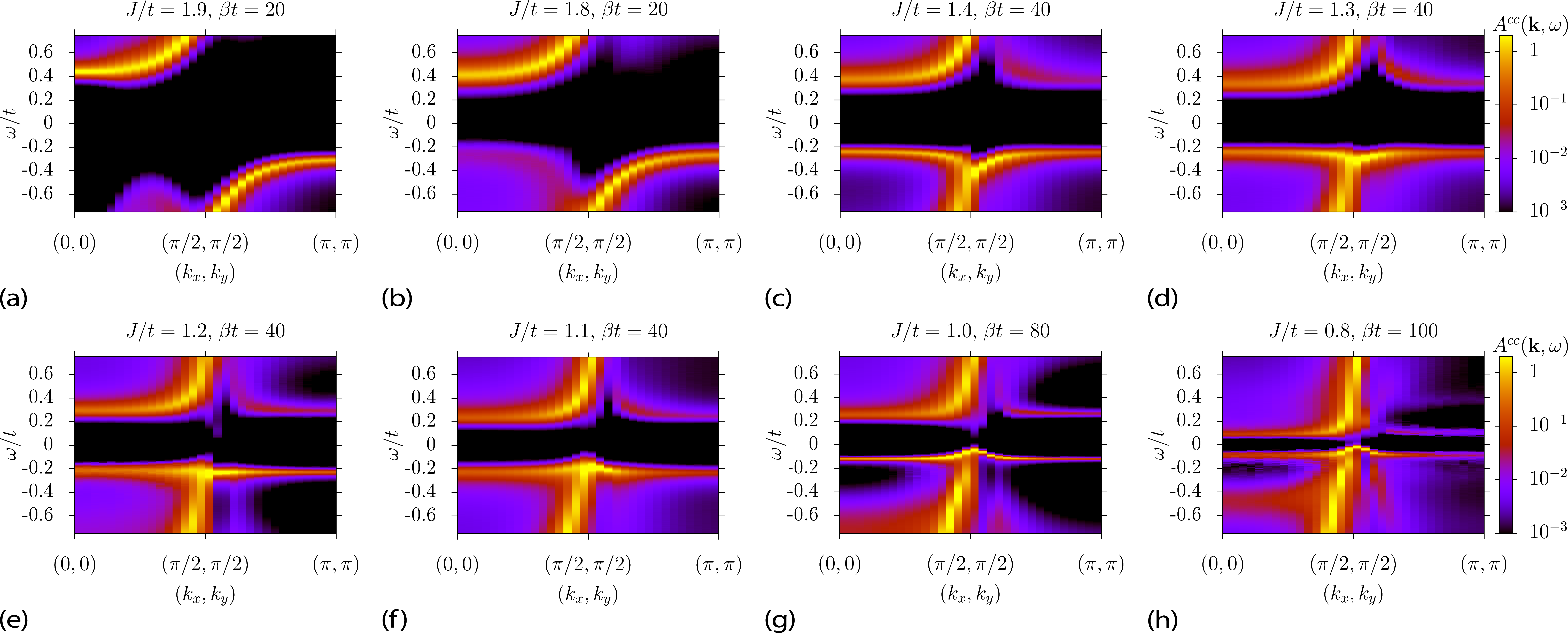}
\caption{(Color online) Closeup for low energies around the Fermi energy (given by $\omega/t =0$). Development of the single-particle spectral function $A^{cc}({\mathbf k}, \omega)$  at half-filling with $t'/t=-0.3$ which shows a shift of the maximum in the lower band from ${\mathbf k}=(\pi,\pi)$ to ${\mathbf k}=(\pi/2,\pi/2)$ with decreasing $J/t$.}
\label{fig:fig7}
\end{center}
\end{figure*}
\begin{figure}
\begin{center}
\includegraphics[width=\columnwidth,type=png,ext=.png,read=.png]{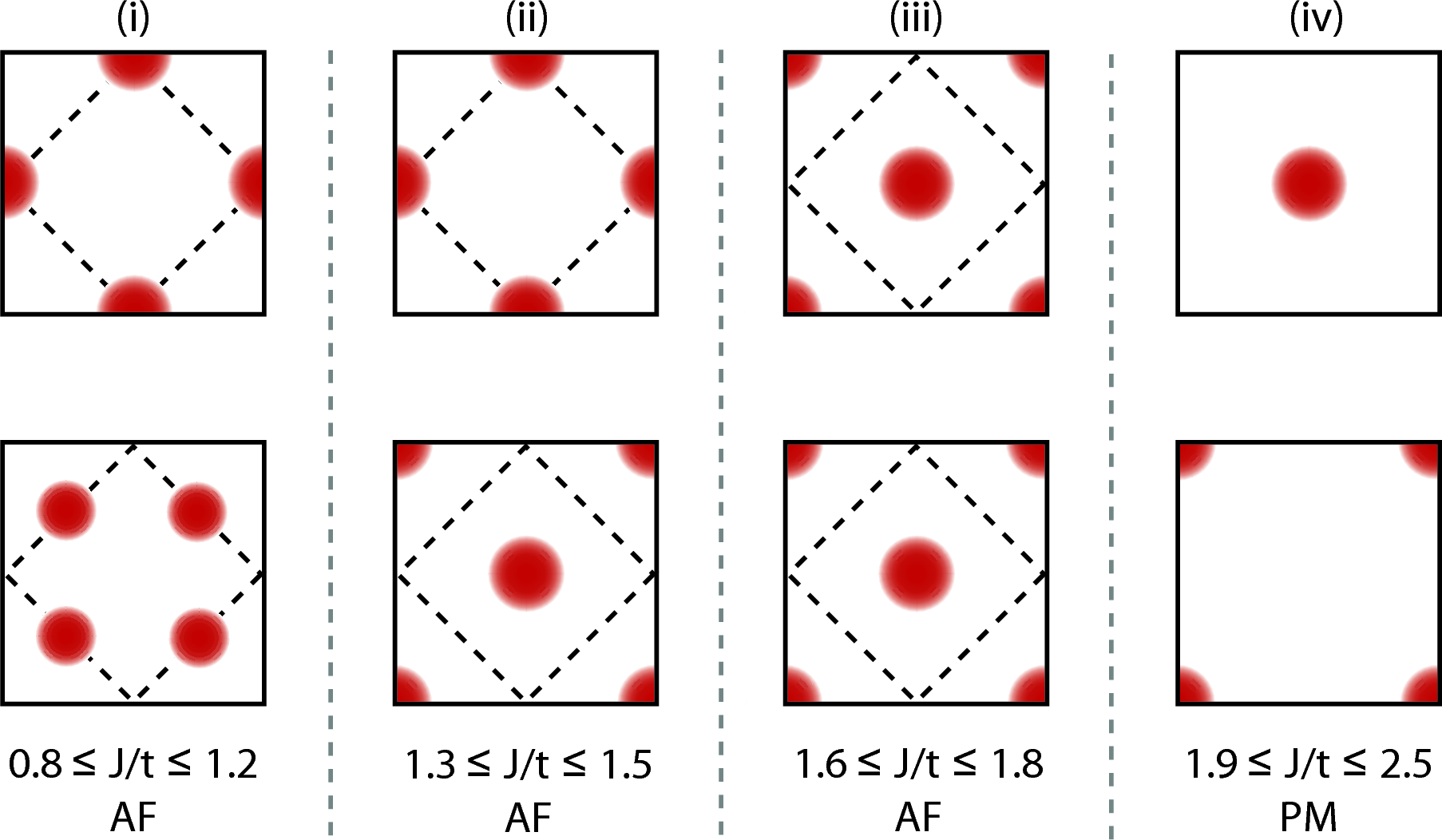} \\
\end{center}
\caption{(Color online) Schematic representation of the position of the upper band minima (top row) and lower band maxima (bottom row) of the conduction electron single-particle spectral function with next-nearest neighbor hopping $t'/t=-0.3$ and for half-filling. The squares represent the first Brillouin zone (for the PM phase) or the extended Brillouin zone (for the AF phase), with the bottom left corner and top right corner of each square given by ${\mathbf k}=(-\pi,-\pi)$ and ${\mathbf k}=(\pi,\pi)$, respectively.}
\label{fig:fig8}
\end{figure}

We now set $t'/t=-0.3$, but remain at half-filling via careful adjustment of the chemical potential $\mu$.  With decreasing $J/t$ we can divide the phase diagram into four regions on the basis of the characteristic spectrum in each region. For $J/t = 2.5$ (Fig.~\ref{fig:fig6}(a)) the spectrum is qualitatively identical to the PM phase in the $t'/t=0$ case such that the model is a Kondo insulator with an indirect gap. The minimum of the valence band lies at ${\mathbf k}=(0,0)$ and the maximum of the conduction  band is at ${\mathbf k}=(\pi,\pi)$.\\
For $J/t=1.8$ (Fig.~\ref{fig:fig6}(b)) the system is weakly AF ordered, and the spectrum reflects this in the development of shadow bands, but is otherwise qualitatively the same as before. At $J/t=1.4$ (Fig.~\ref{fig:fig6}(c)), well inside the AF phase, the shadow bands are more pronounced and now the minimum of the valence band has shifted to ${\mathbf k}=(0,\pi)$. The final plot in the series (Fig.~\ref{fig:fig6}(d)), with $J/t=1.0$ shows that the form of the lower band has also changed such that the maximum of this band now lies at ${\mathbf k}=(\pi/2,\pi/2)$.\\
In Figs.~\ref{fig:fig7}(a-h) we show the evolution of this lower band between ${\mathbf k}=(0,0)$ and ${\mathbf k}=(\pi,\pi)$ with an enlargement of the energy axis around the Fermi energy.
The local minimum energy dip at ${\mathbf k}=(\pi/2,\pi/2)$ becomes less pronounced as the heavy-fermion band flattens until by the time we reach $J/t=1.2$ this dip has turned into a bump such that the lower band maximum has shifted from ${\mathbf k}=(\pi,\pi)$ to ${\mathbf k}=(\pi/2,\pi/2)$. This important change in band structure appears to be continuous. For smaller coupling, $J/t=1.1$, $1.0$, and $0.8$ the lower band maximum remains at ${\mathbf k}=(\pi/2,\pi/2)$ and becomes more pronounced.  
Assuming a rigid  band picture, this  evolution of the band structure maps onto a topology change of the Fermi surface  at small dopings away from half filling.  We will see by explicit calculations at finite dopings that this topology change indeed  occurs.\\
The position of the minima and maxima in the valence and conduction  bands of the spectra for $t'/t=-0.3$ are summarized 
schematically in Fig.~\ref{fig:fig8}. 
\subsection{The quasi-particle gap}
\label{sec:subsection3C}
The quasi-particle gap  corresponds to  the energy difference  between the top of the conduction band and the bottom of the valence band.   To be more precise,
\begin{equation}
	2 \Delta_{qp}  =   \min_{{\mathbf k}} \left( E_0^{N+1}({\mathbf k}) - E_0^N \right)   - 
                           \max_{{\mathbf k}} \left( E_0^N -  E_0^{N-1}({\mathbf k}) \right). 
\label{eqn:eqn7}
\end{equation}
Here,  $ E_0^{N \pm 1}({\mathbf k}) $   corresponds to the ground state energy   of the $N \pm 1$ particle system with total 
momentum  ${\mathbf k}$ and $ E_0^N  =  \min_{{\mathbf k}} E_0^N({\mathbf k}) $.   Technically, the energy 
differences   are at best extracted from  the low temperature single-particle Green functions:
\begin{align*}
	& \lim_{ \beta \rightarrow \infty } \langle c(\tau)_{{\mathbf k}} c^{\dagger}_{{\mathbf k}} \rangle   \propto 
	e^{ -\tau (E_0^{N+1}({\mathbf k})  -  E_0^{N}     - \mu )  }   \; \;,     \tau >> 1  \\
        & \lim_{ \beta \rightarrow \infty } \langle c^{\dagger}_{{\mathbf k}}(\tau) c_{{\mathbf k}} \rangle   \propto 
	e^{  \tau ( E_0^{N}    - E_0^{N-1}({\mathbf k})  - \mu ) }    \; \;,     \tau << -1\;.
\end{align*}
\begin{figure}
\begin{center}
\includegraphics[width=\columnwidth,type=png,ext=.png,read=.png]{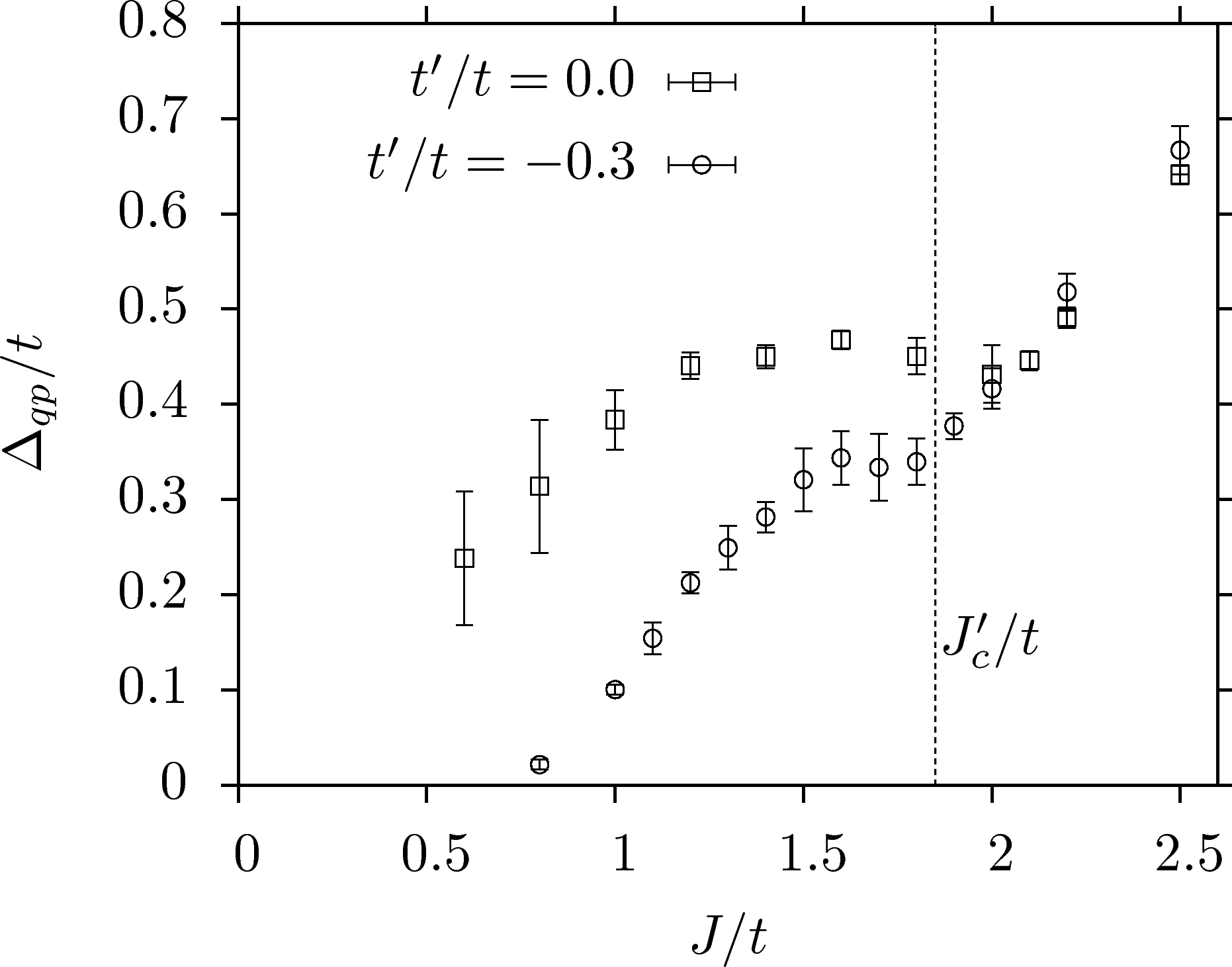}
\end{center}
\caption{The quasi-particle gap $\Delta_{qp}/t$ as a function of Kondo coupling $J/t$. At $t' = -0.3 t$, the magnetic order-disorder transition takes place at $J'_{c}/t \simeq 1.85$. In the coupling range  $ 0.8  < J/t < J'_{c}/t$ Kondo screening coexists with magnetism and is at the origin of the quasiparticle gap.}
\label{fig:fig9}
\end{figure} 
In the large $J/t$ limit, each impurity spin  traps a conduction electron in a Kondo singlet. The wave function corresponds to a direct product of such Kondo  singlets and the quasi-particle gap is set by the energy scale $4J/3$ required to break  a Kondo singlet.  Hence, in the large $J/t$ limit the quasi-particle gap  does not depend on the details of the band 
structure.   The situation is more subtle in the  weak-coupling limit, where the underlying nesting properties of the  Fermi surface play a crucial role.  Assuming static impurity spins locked into  an antiferromagnetic order, 
$ \langle {\mathbf S}_{{\mathbf i}}^{f} \rangle = \frac{1}{4} {\mathbf e}_z e ^ {i {\mathbf Q} \cdot {\mathbf i}} $
one obtains the single particle 
dispersion relation
\begin{equation}
	E_{\pm}({\mathbf k} ) = \epsilon_{+}({\mathbf k}) \pm 
   \sqrt{ \epsilon_{-}^{2}({\mathbf k})  + (J/4)^{2} }\;,
\label{eqn:eqn8}
\end{equation}
with  $\epsilon_{\pm}({\mathbf k})  =  \frac{ \epsilon({\mathbf k}) \pm \epsilon({\mathbf k}+{\mathbf Q} ) }{2}   $.
In one dimension and at $t'/t=0$   nesting leads to $\epsilon_{+}({\mathbf k}) \equiv 0 $  and  magnetic order opens a 
quasi-particle  gap set by  $\Delta_{qp} = J/4$.   Both the DCA results presented in Fig. \ref{fig:fig9} 
as well as  the BSS results of Ref. \onlinecite{Capponi00} support this point of view.   In one-dimension,  this scaling of the 
quasi-particle gap is also observed \cite{Tsunetsugu97_rev}. 
At $t'/t = -0.3$  nesting is not present and  magnetic ordering can only partially gap the Fermi surface.  
Since the DCA results of Fig. (\ref{fig:fig9})   support a finite quasi-particle gap  in the magnetically ordered phase, 
 the above frozen-spin Ansatz fails and points to one of the main  result of our work, namely that within the 
magnetically ordered phase Kondo screening and the heavy quasi-particles are still present.  
Even though we cannot track the functional form of the gap at $t'/t = -0.3 $ and at weak couplings,  we interpret the DCA results in terms of a quasi-particle gap which tracks the Kondo scale from weak to strong coupling as in the large-$N$ mean-field calculations \cite{Georges00}.
\section{The Hole-Doped KLM}
\label{sec:section4}
In this section we concentrate on the hole-doped KLM at $t'/t = -0.3$. We will first map out  the magnetic phase diagram in the doping versus coupling  plane and then study the evolution of  single-particle spectral function  from the previously discussed half-filled case   to the heavily doped paramagnetic heavy-fermion metallic  state.   
\subsection{Magnetic phase diagram}
\label{sec:subsection4A}
\begin{figure}
\begin{center}
\includegraphics[width=\columnwidth,type=png,ext=.png,read=.png]{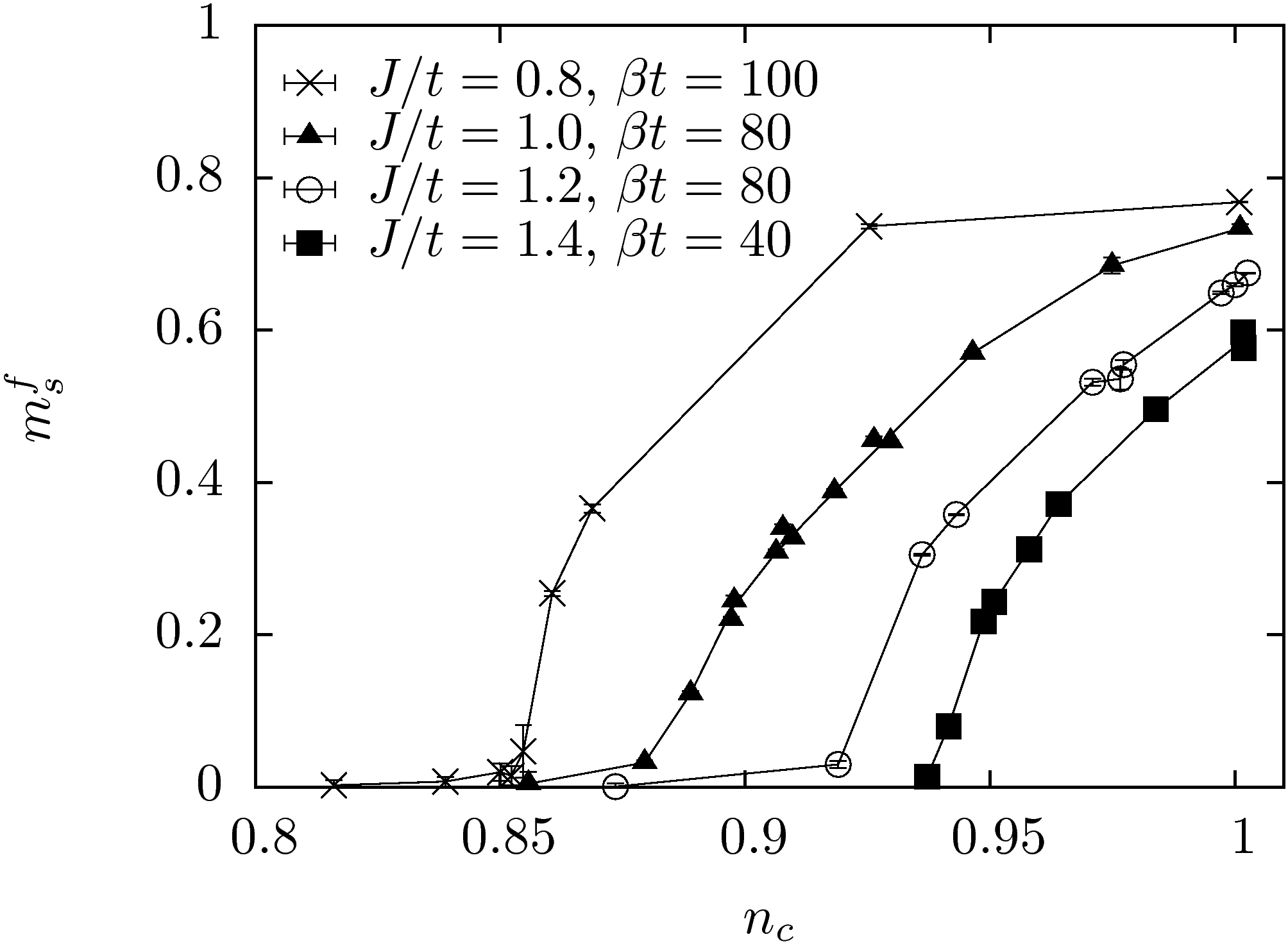}
\caption{The staggered magnetization $m^{f}_{s}$ of the $f$-electrons as a function of $n_{c}$ at different constant couplings $J/t$.}
\label{fig:fig10}
\end{center}
\end{figure}
\begin{figure}
\begin{center}
\includegraphics[width=\columnwidth,type=png,ext=.png,read=.png]{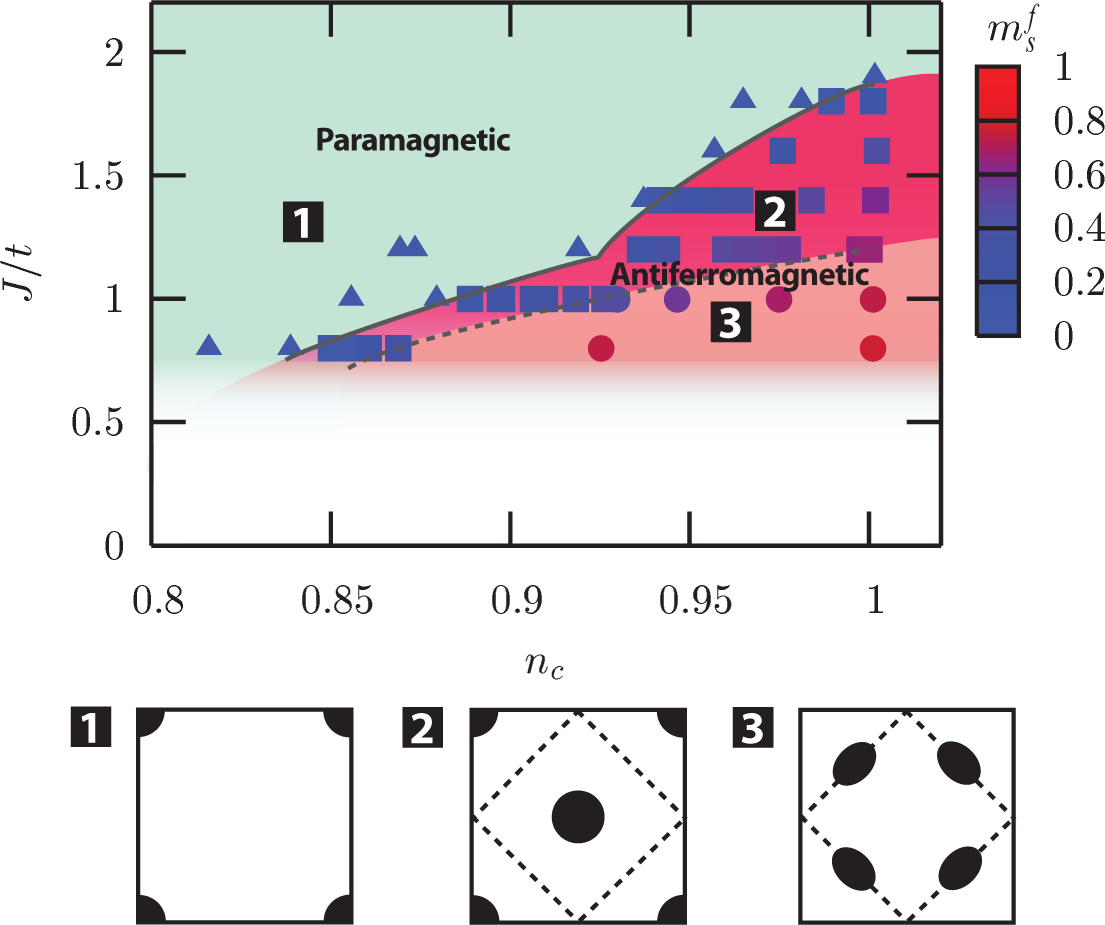}  \\
\end{center}
\caption{(Color online) Ground state magnetic phase diagram of the hole-doped KLM showing simulation results for the staggered magnetization $m^{f}_{s}$ (color-coded) as a function of coupling $J/t$ and conduction electron occupancy $n_{c}$. Triangles: PM region, large FS. Squares: AF, large FS. Circles: AF, small FS. Here $t'/t = -0.3$ and the calculations are carried out with the $N_{p} = 1$ cluster. Below, the FS topologies corresponding to the numbered regions are shown schematically.}
\label{fig:fig11}
\end{figure}

Fig.~\ref{fig:fig10} shows the staggered magnetization as a function of conduction electron density for $J/t=0.8$, $1.0$, $1.2$ and $1.4$. The magnetically ordered state found at half-filling initially survives when doping with holes. At all coupling values a continuous magnetic phase transition is observed with the AF order decreasing gradually as the system is doped and vanishing smoothly at a quantum critical point.   The results are summarized in the magnetic phase diagram shown in Fig. \ref{fig:fig11}. With decreasing values of $J/t$ and increasing dopings the RKKY interaction progressively dominates over the Kondo scale  and the magnetic metallic state is  stabilized. Checks were made by varying the temperature, $1/\beta$, of the simulations to ensure that the results can be considered to be ground state.   Below $J/t = 0.8$  we  found the coherence scale to be too low to  guarantee convergence. As pointed out previously,  the limiting computational factor is the cubic scaling of the Hirsch-Fye \cite{HirschFye86} algorithm,   $(\beta N_{p})^3$ and not the negative sign problem. 
\subsection{Single-particle spectrum and topology of the Fermi surface}
\label{sec:subsection4B}
\begin{figure*}
\begin{center}
\includegraphics[width=\textwidth,type=png,ext=.png,read=.png]{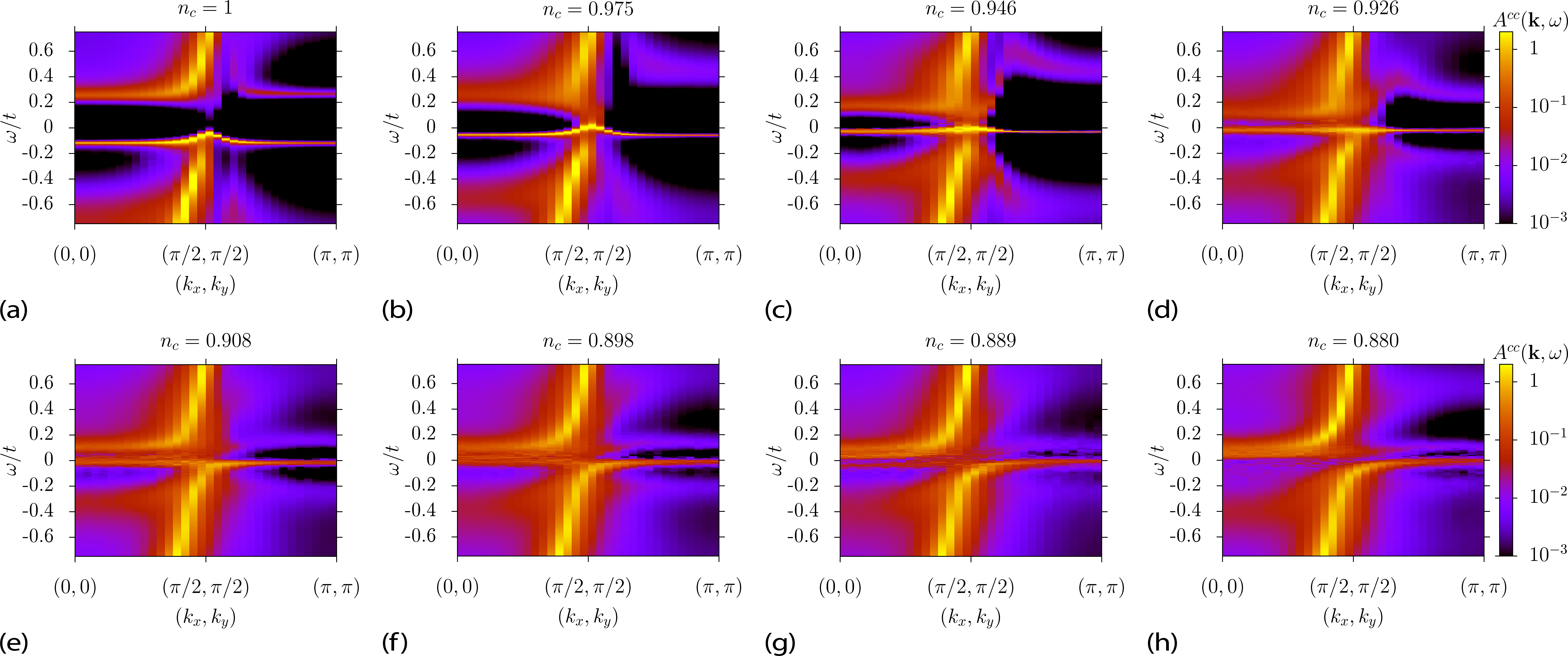}
\end{center}
\caption{(Color online) Single-particle spectra $A^{cc}({\mathbf k}, \omega)$ of the conduction electrons for $J/t=1$ and $t'/t=-0.3$, at $\beta t = 80$. The conduction electron density is reduced progressively from half-filling $n_{c} = 1$ (a)  to $n_{c} =  0.880$ (h). The respective FS topologies are shown in Fig.~\ref{fig:fig11}.}
\label{fig:fig12}
\end{figure*}

We have calculated and followed the evolution of the single-particle spectrum, 
plotted in an extended Brillouin zone scheme.
To begin with we set $J/t=1$ and show results in Fig.~\ref{fig:fig12} for the spectral function,  
as it is here that a change in the Fermi surface topology becomes evident.  
Starting from the previously discussed half-filled case  (Fig.~\ref{fig:fig12}(a)) a rigid band approximation produces hole pockets around the ${\mathbf k}=(\pm \pi/2, \pm \pi/2) $ points in the Brillouin zone. This  is confirmed by explicit calculations at  small dopings 
away from half-filling  as demonstrated in Fig.~\ref{fig:fig12}(b). 
With further doping this low energy band flattens out progressively becoming almost flat by $n_{c} = 0.926$ (Fig.~\ref{fig:fig12}(d)). Doping further  gives rise to a  Fermi surface with holes centered around  ${\mathbf k}=(\pm \pi, \pm \pi)$. Since at $n_{c} = 0.908$ (Fig.~\ref{fig:fig12}(e)) and $n_{c} = 0.898$ (Fig.~\ref{fig:fig12}(f)) we still have non-zero magnetizations of $m^{f}_{s}=0.340$ and $m^{f}_{s}=0.245$, respectively, shadow features centered around  ${\mathbf k}=(0,0)$ are expected. The weight of those shadow features progressively diminishes as the staggered magnetization vanishes.  In particular, in Fig.~\ref{fig:fig12}(g) the magnetization is very small ($m^{f}_{s}=0.123$)  and 
indistinguishable  from zero in the final figure of the series.\\
Across the transition from $n_{c}=1$ to $n_{c}=0.856$ the quasiparticle weight 
$Z({\mathbf k}=(\pi,\pi))=|\langle \Psi_{0}^{N+1}|c_{{\mathbf k}=(\pi,\pi),\sigma}^{\dagger}|\Psi_{0}^{N}\rangle|^{2}$
 as obtained from the behaviour of $g^{cc}({\mathbf k}=(\pi,\pi),\tau)$ at large imaginary times decreases albeit it shows no sign of singularity. This observation stands in agreement  with results obtained at half-band filling \cite {Assaad04a} and excludes  the occurrence of a Kondo breakdown.
\\
This evolution of the single-particle spectral function at $J/t = 1$  points to three distinct Fermi surface topologies, sketched in 
Fig. \ref{fig:fig11}.    In the paramagnetic phase (Fermi surface (1) in Fig. \ref{fig:fig11}) the Fermi surface  consists of hole pockets around the ${\mathbf k}=(\pm \pi, \pm \pi) $ points in the  Brillouin zone.   Even though in our simulations  charge fluctuations of the $f$-sites are completely prohibited,   the Fermi surface volume accounts for both the conduction  electrons and impurity spins. 
This Fermi surface topology maps onto that of the corresponding  non-interacting periodic Anderson model with total particle density given by $1 + n_c$ and is coined large Fermi surface.    In the antiferromagnetic metallic phase close to the magnetic order-disorder transition,  (Fermi surface (2) in  Fig. \ref{fig:fig11}) the Fermi surface merely corresponds to a  backfolding of the paramagnetic Fermi surface  as expected in a generic spin-density wave transition.   Here,  a heavy  quasi-particle  with crystal momentum ${\mathbf k}$ can scatter  off a  magnon with momentum  ${\mathbf Q} = (\pi,\pi) $ to produce a shadow feature at ${\mathbf k}+{\mathbf Q}$.     
It is only  within the magnetically ordered phase that we  observe the topology change of the Fermi  surface to 
hole pockets centered around $ {\mathbf k}= ( \pm \pi/2, \pm \pi/2)$.   Due to the antiferromagnetic order and the accompanied 
reduced Brillouin zone, this  Fermi surface topology   satisfies the Luttinger sum rule. We  coin it a small  Fermi surface in the 
sense that it is linked to that obtained in a frozen $f$-spin  mean-field calculation as presented 
in Eq. (\ref{eqn:eqn9}).
\begin{figure}
\begin{center}
\includegraphics[width=\columnwidth,type=png,ext=.png,read=.png]{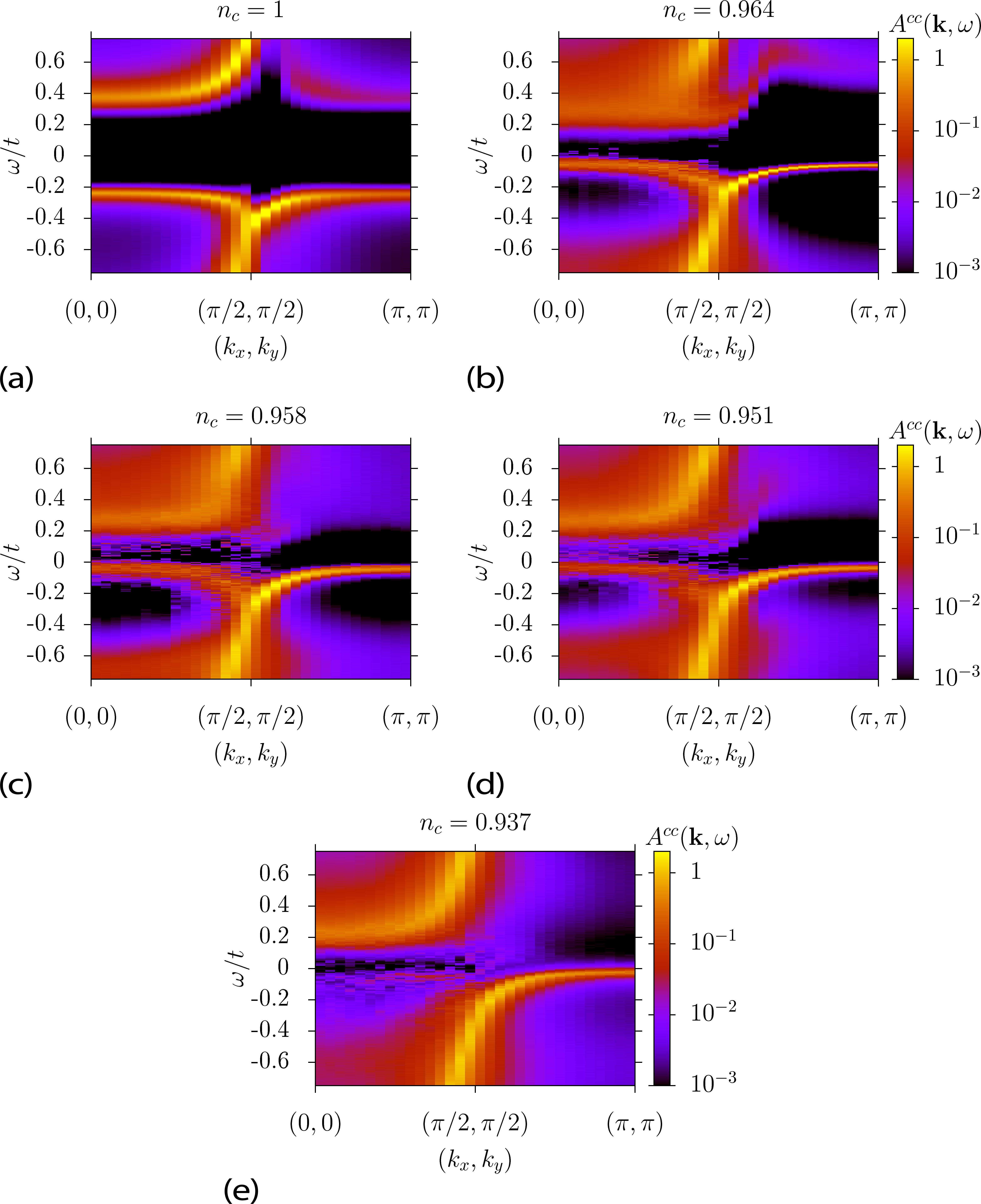}
\end{center}
\caption{(Color online) Single-particle spectra $A^{cc}({\mathbf k}, \omega)$ of the conduction electrons for $J/t=1.4$ and $t'/t=-0.3$, at $\beta t = 40$. The shadow features present in (a) vanish as the occupation number $n_{c}$ is reduced.}
\label{fig:fig13}
\end{figure}
We have equally  plotted the  single-particle spectral  function at higher values of $J/t=1.4$ as a function of doping 
(See Fig. \ref{fig:fig13}). For this coupling strength, and at half-band filling, the conduction band  maximum is located 
at ${\mathbf k}=(\pi,\pi)$  with accompanying shadow bands at ${\mathbf k}=(0,0)$. The Fermi surface obtained upon doping  follows  from a rigid band approximation, and yields  the  topology  shown in Fig. \ref{fig:fig11} (2).  The transition to the paramagnetic state   shows up in the vanishing of the shadow features at $ {\mathbf k}= (0,0) $.\\
We note that our results are confirmed by simulations carried out using the larger 
cluster size $N_{p}=4$, see Fig.~\ref{fig:fig14}.
 In particular, we observe the small Fermi surface topology in the strong AF region with small coupling $J/t$ and evidence for a large Fermi surface in the weakly ordered region ($J/t \gtrsim 1.2$). 
\begin{figure}
\begin{center}
\includegraphics[width=\columnwidth,type=png,ext=.png,read=.png]{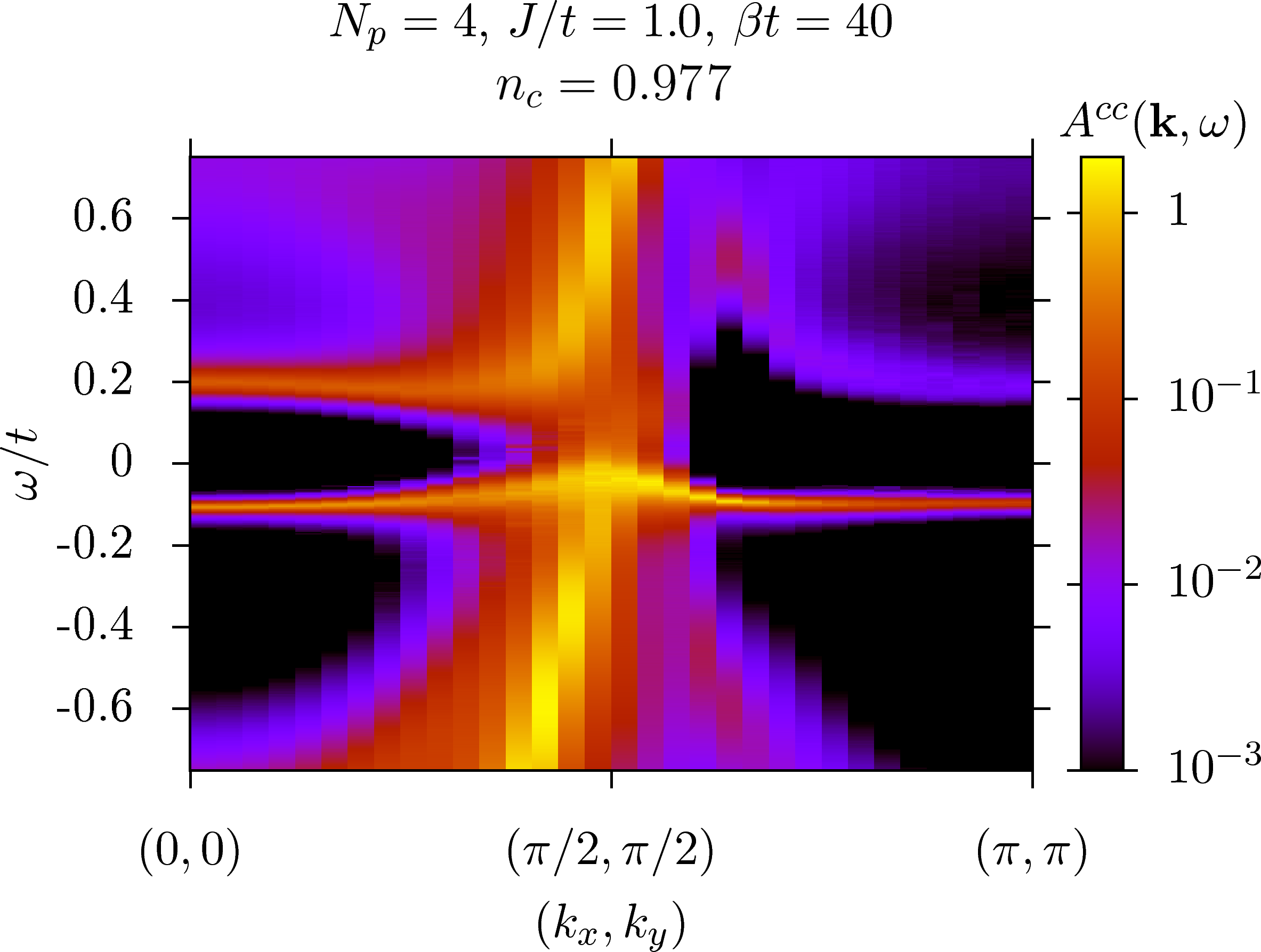}
\end{center}
\caption{(Color online) 
Single-particle spectrum $A^{cc}({\mathbf k}, \omega)$ of the conduction electrons obtained using cluster size $N_{p}=4$. confirming that the small Fermi surface topology in the lightly doped AF phase with coupling $J/t=1$ 
is not an artifact of the smaller cluster results.  The staggered magnetisation of the $f$-electrons is  $m_{s}^{f} = 0.657 \pm 0.005$.}
\label{fig:fig14}
\end{figure}
\subsection{Mean-field modeling of the single-particle spectral function}
\label{sec:subsection4C}
\begin{figure}
\begin{center} 
\includegraphics[width=\columnwidth,type=png,ext=.png,read=.png]{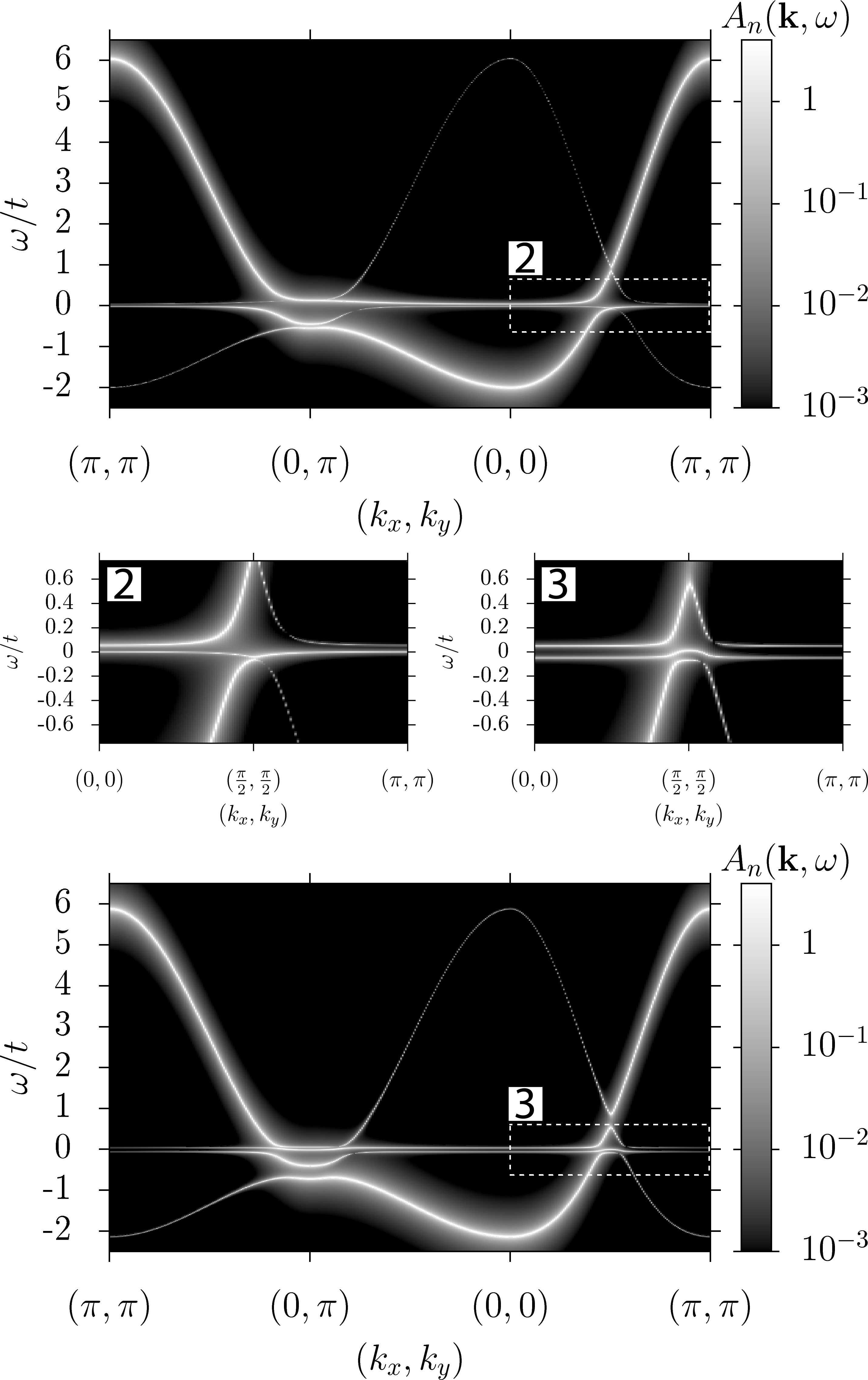}
\caption{  Mean-field band structures $E_{n}({\mathbf k})$ with $J/t=1$ ($t'/t=-0.3$). (a) $V = 0.5, m^{c}_{s}=0.066,m^{f}_{s}=0.257$, $n_{c}=0.898$. The line intensity is proportional to the quasi-particle spectral weight, $A_{n}({\mathbf k}, \omega)$ . The Fermi surface topology (see closeup (b)) results from backfolding of the large-$N$ mean-field  bands.  This corresponds to  the Fermi surface (2) in Fig.\ref{fig:fig11}.\\
(d) $V = 0.260, m^{c}_{s}=0.190,m^{f}_{s}=0.693$, $n_{c}=0.975$. The Fermi surface topology (see closeup (c)) corresponds to that of hole pockets centered around ${\mathbf k}=( \pm \pi/2, \pm \pi/2 )$ as depicted  by the Fermi surface (3) in Fig. \ref{fig:fig11}.}
\label{fig:fig15}
\end{center}
\end{figure}

\begin{figure}
\begin{center}
\includegraphics[width=\columnwidth,type=png,ext=.png,read=.png]{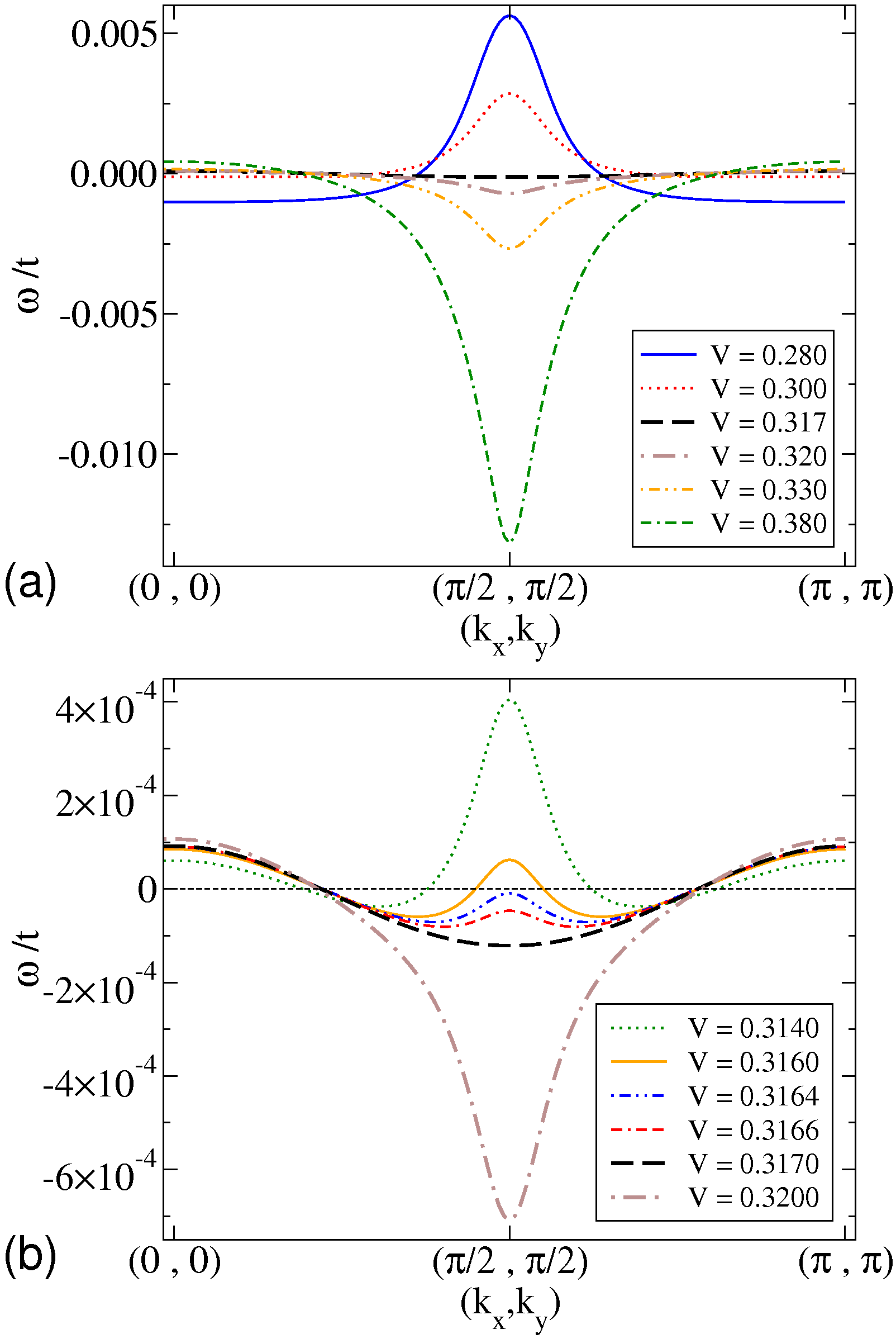}
\caption{(Color online) Topology change (a) of the Fermi surface as obtained from the mean-field modeling $E_{n}({\mathbf k})$  of the DCA results ($J/t=1$, $t'/t=-0.3$ $m^{c}_{s}=0.066$, $m^{f}_{s}=0.257$, $n_{c}=0.898$).
As apparent on the scale of the closeup (b), hole pockets around ${\mathbf k}=(\pm \pi/2,\pm \pi/2) $ as well as 
around ${\mathbf k}=(\pi,\pi) $  are present at $V=0.314$.}
\label{fig:fig16}
\end{center}
\end{figure}

We resort to a mean-field modelling in order to propose a scenario for the detailed nature of the topological transition of the FS within the ordered phase. Within this model the topology change will correspond to two Lifshitz transitions.
Aspects of the  single-particle spectral function can be well accounted for within the following mean-field Hamiltonian \cite{Zhang00b}:
\begin{eqnarray}
\tilde{H}&=&\sum_{{\mathbf k} \sigma} 
\left(
\begin{array}{c}
c_{{\mathbf k} \sigma} \\
c_{{\mathbf k} + {\mathbf Q} \sigma} \\
f_{{\mathbf k} \sigma}\\
f_{{\mathbf k} + {\mathbf Q} \sigma}
\end{array}
\right)^{\dagger} \nonumber \\
& &\times \left(
\begin{array}{cccc}
\epsilon_{{\mathbf k}} -\mu     & \frac{J m^{f}_{s} \sigma}{4}        & \frac{J V}{2}               & 0                           \\
\frac{J m^{f}_{s} \sigma}{4}   & \epsilon_{{\mathbf k} + {\mathbf Q}} -\mu &       0                     & \frac{J V}{2}               \\
\frac{J V}{2}              &          0                      &   \lambda                   & -\frac{J m^{c}_{s} \sigma}{4}   \\
         0                 &       \frac{J V}{2}             &  -\frac{J m^{c}_{s} \sigma}{4}  &   \lambda
\end{array}
\right) \nonumber \\
& &\times \left(
\begin{array}{c}
c_{{\mathbf k} \sigma} \\
c_{{\mathbf k} + {\mathbf Q} \sigma} \\
f_{{\mathbf k} \sigma}\\
f_{{\mathbf k} + {\mathbf Q} \sigma}
\end{array}
\right)\;.
\label{eqn:eqn9}
\end{eqnarray}
Here $ \lambda $, ($\mu$) are Lagrange multipliers fixing the $f$- ($c$-) particle  number to unity ($n_{c}$) and  $m^{c}_{s}$,  $m^{f}_{s}$  
corresponds to the staggered magnetization of the conduction and   $f$-electrons. The order parameter for Kondo screening is $V\propto\langle  f^{\dagger}_{i, \sigma} c_{i, \sigma } + c^{\dagger}_{i, -\sigma} f_{i, -\sigma }  \rangle$. The  large-$N$ mean-field saddle point 
corresponds to the choice $m^{c}_{s}=m^{f}_{s}=0$ and $V  \neq 0$. As already mentioned this saddle point  gives a  good account of the 
hybridized bands we observe numerically  in the paramagnetic phase.  In the magnetic phase, one could speculate that Kondo 
screening breaks down such  that  the $f$-spins are frozen and do not participate in the Luttinger volume. This corresponds to 
the  parameter set $V = 0$, $m^{c}_{s} \neq 0$ and $m^{f}_{s} \neq 0$ and leads to the dispersion relation of Eq. (\ref{eqn:eqn8}). 
Throughout the phase diagram of Fig. \ref{fig:fig11}  the single-particle spectral function never shows features following this scenario.   In fact,  many aspects  of the spectral function in the magnetically ordered phase can be 
understood by choosing   $m^{c}_{s}\neq 0$,  $m^{f}_{s}\neq 0$ and $V \neq 0$. This  explicitly accounts for a heavy-fermion band in the ordered phase. It is in this sense that we claim the absence of Kondo breakdown  within our model calculations.  
Fig. \ref{fig:fig15}  plots the four-band energy dispersion relation $E_{n}({\mathbf k})$ obtained from the mean-field calculation in the magnetically ordered phase at 
$t'/t=-0.3$, $J/t=1$  and  at band fillings $n_{c}=0.898$ (Fig.~\ref{fig:fig15}(a,b))   and $n_{c}=0.975$ (Fig.~\ref{fig:fig15}(c,d)) corresponding to the  Fermi surface topologies labeled (2) and (3), respectively, in  
Fig. \ref{fig:fig11}. 
In each case we have used the  DCA value for the magnetization and  varied $V$ to obtain the best qualitative fit to the respective DCA spectrum.   The change in the topology of the Fermi surface  stems from a delicate interplay  of the relative magnitudes of the magnetization  $m^{c,f}_{s}$ and 
{\it Kondo screening } parameter $V$.\\
In the DCA calculations, we are unable to study the precise  nature  of the topology  change of the Fermi surface   within the magnetically ordered phase since the energy scales are too small. However we can draw on the mean-field model to gain some insight. Fig. \ref{fig:fig16}  plots the dispersion relation along the  ${\mathbf k}=(0,0)$ to ${\mathbf k}=(\pi,\pi)$  line in the  Brillouin zone as a function of $V$.   We would like to emphasize the following  points. i) The transition from hole pockets  around the wave vector ${\mathbf k}=(\pi,\pi) $ to 
${\mathbf k}=(\pm \pi/2,\pm \pi/2)$  corresponds to two Lifshitz transitions in which the hole pockets around  
${\mathbf k}=(\pi,\pi) $   disappears and those around ${\mathbf k}=(\pm \pi/2,\pm \pi/2) $  emerge. 
The Luttinger sum rule requires an intermediate phase with the presence of  pockets  both at  ${\mathbf k}=(\pm \pi/2,\pm \pi/2) $ and at ${\mathbf k}=(\pi,\pi) $. This situation is explicitly seen in
Fig.~\ref{fig:fig16}(b) at $V = 0.314$.  ii) Starting from the antiferromagnetic  state with 
hole pockets at ${\mathbf k}=( \pm \pi/2, \pm \pi/2 )$  there is an overall flatting of the band  prior to the  change in the Fermi surface topology.  Hence on both sides of the transition,  an enhanced effective mass is expected. Owing to point i)  the effective mass does not diverge. 
iii)   The two Lifshitz transition scenario suggested by this mean-field modeling,  
leads to a continuous  change of the Hall 
coefficient.  However, since a very {\it small}  variation of  the hybridization $V$ suffices to change the Fermi 
surface topology  one can  expect, as a function of this control parameter, a rapid variation  of the Hall coefficient as it is observed in experiment \cite{Paschen04}.
\subsection{Discussion}
\label{sec:subsection4D}
The evolution of the Fermi surface  across the magnetic order-disorder transition has  been investigated by other groups and especially in the framework of  a Gutzwiller projected mean-field wave function  corresponding to  the ground state of  the  single particle Hamiltonian of Eq. \ref{eqn:eqn9}. Both a  variational quantum Monte Carlo calculation  \cite{Watanabe07,Watanabe09} as well as a   Gutzwiller approximation \cite{Lanata08}  show  a rich phase diagram which bears some similarities but also important differences with the present DCA calculation. At small doping away from half-filling, the magnetic transition  as a function of the coupling $J/t$ is continuous  and of SDW type  as marked by the back-folding of  the Fermi surface.  Within the magnetically ordered phase a first order transition  occurs  between  hole (AF$_h$)  and electron (AF$_e$)   like  Fermi surfaces.  At larger hole dopings,    the magnetic transition is of  first order  and the  Fermi surface  abruptly changes from large to small (AF$_e$).  The calculations in  Refs.~\onlinecite{Watanabe07,Watanabe09,Lanata08}  are carried out at $t' =0$. At finite values of $t'= -0.3t$  one expects the  AF$_e$ Fermi-surface topology to correspond to hole pockets centered around   $\vec{k} = (\pm \pi/2, \pm \pi/2)$ as shown in Fig.~\ref{fig:fig12}.  On the other hand, the  AF$_h$ Fermi surface arises from a back-folding of the large Fermi surface (Fig.~\ref{fig:fig13}).  With this identification, our  DCA results bear some  similarity with the variational calculation  in the sense that the same  Fermi surface topologies are realized.  However, in our calculations we see no sign of  first order transitions and no direct transitions from the large paramagnetic Fermi-surface topology to the AF$_e$ phase. It is equally important to note that in both the variational and DCA  approaches no Kondo breakdown is apparent.   In particular the variational parameter $ \tilde{V}$ which encodes hybridization between the $f$- and $c$-electrons  never vanishes  \cite{Watanabe07,Watanabe09}.  
On the other hand, $T=0$ CDMFT calculations of the three-dimensional periodic Anderson model on a two site cluster and finite sized baths \cite{DeLeo08,DeLeo08a} have put forward the idea that the magnetic order-disorder transition is driven by an orbital selective Mott transition  or in other words a Kondo breakdown \cite{Vojta10}. In particular even in the presence of spin symmetry broken baths allowing for antiferromagnetic ordering a big  enhancement of the $f$-effective mass is apparent in the vicinity of the magnetic transition.    In the paramagnetic phase,  the low energy decoupling of the $f$- and $c$-electrons stem from the vanishing of the hybridization function at low energies.
Our results,  which makes no approximation of the  number of bath degrees of freedom and which allow to treat larger cluster  sizes do not support this point of view. In particular as shown in Ref.~\onlinecite{Assaad04a} no singularity in the quasiparticle weight  is apparent across the transition. Equally, Fig. \ref{fig:fig12} which plots the single particle spectral function at finite doping shows no singularity in $Z({\mathbf k}=(\pi,\pi))$ .
\section{Conclusion}
\label{sec:section5}
We have presented detailed  DCA+QMC  results  for the Kondo Lattice model on a square lattice.   Our DCA approach allows for antiferromagnetic order such that  spectral functions can be computed  across magnetic transitions and  in magnetically ordered phases. At the particle-hole symmetric point (half-band filling and $t'=0$)  we can compare this approach to 
previous  {\it exact} lattice QMC  auxiliary field simulations \cite{Capponi00,Assaad08_rev}.   Unsurprisingly, with the {\it small} cluster sizes considered in the DCA approach the 
critical  value of $J/t$ at which the  magnetic order-disorder transition occurs is substantially overestimated.  The important point however is  that the DCA+QMC approximation  gives an extremely good account of the single-particle spectral function both in the paramagnetic  and antiferromagnetic phases.   
Hence, and as far as we can test against benchmark results, the  combination of static magnetic ordering and  dynamical Kondo  screening, as realized in the DCA,   provides a very good  approximation of the underlying physics.\\
As opposed to {\it exact} lattice QMC  auxiliary field simulations, which fail  away from the particle-hole symmetric point 
due to the  so-called negative sign problem, the DCA+QMC approach allows us to take steps away from this symmetry either by doping or by introducing a finite value of $t'$.  It is worthwhile noting that for cluster sizes up to $16$ orbitals the limiting factor  is the cubic scaling of the Hirsch-Fye algorithm rather than the negative sign problem which   turns out to be very mild in the considered parameter range.\\
Our major findings are summarized   by the following points. \\
i)  We observe no Kondo breakdown throughout the phase diagram even  deep in the  antiferromagnetic ordered 
state where the staggered magnetization takes large values.  This  has the most dramatic effect at half-band filling 
away from the particle-hole symmetric point and at small values of $J/t$ where  magnetic order is robust. 
Here, magnetism alone will not account for the  insulating state and the  observed quasi-particle gap can only be 
interpreted in terms  of  Kondo screening.    The absence of Kondo breakdown  throughout the phase diagram 
is  equally confirmed by the single-particle spectral function which always shows a feature reminiscent of the 
heavy-fermion band.\cite{footnote}  \\
ii) The transition  from the paramagnetic to the antiferromagnetic  state is continuous and is associated  
with the backfolding of the  large heavy fermion Fermi surface.  \\
iii)   Within the antiferromagnetic metallic phase there is a Fermi surface topology change from hole 
pockets centered around  ${\mathbf k}=(\pi,\pi)$  at {\it small} values of the magnetization $m^{f}_{s}$ to one centered  
around  ${\mathbf k}=(\pi/2,\pi/2) $ at {\it larger} values of $m^{f}_{s}$. The latter Fermi surface is adiabatically 
connected to one where the $f$-spins are frozen  and in which  Kondo screening is completely absent. \cite{Vojta08, Vojta10}   This transition comes about by {\it continuously deforming } a heavy-fermion band  such that the effective mass grows substantially on  both sides of the transition.   Within a mean-field modelling of the quantum Monte Carlo results  the transition  in the Fermi surface topology corresponds to two continuous Lifshitz transitions. As a function of decreasing mean-field Kondo screening parameter $V$ and at constant  hole doping,  the volume of the hole pocket 
at ${\mathbf k}=(\pi,\pi)$ decreases at the expense of the increase of volume of the hole pocket at ${\mathbf k}=(\pi/2,\pi/2)$.  Owing to 
the Luttinger sum rule, the  total volume of the  hole pockets remains constant. It is important to note that the energy scales required to resolve  the  nature of this transition are presently orders of magnitudes smaller than accessible  within the DCA with a quantum Monte Carlo solver.\\
\\
\acknowledgments
We would like to thank the DFG for financial support under the grant numbers  AS120/6-1 (FOR1162)  and  AS120/4-3.  The numerical calculations were carried out at the LRZ-Munich as well as at the JSC-J\"ulich.  We thank those institutions for generous allocation of CPU time. 
\bibliographystyle{./apsrev4-1}

\begin{thebibliography}{53}
\bibitem[{\citenamefont{}()}]{footnote}
  \bibinfo{}{
A scenario where  Kondo breakdown occurs can be deliberately induced by
an energy scale that corresponds to the zero-point fluctuation of the
$f$-spins. In our Hamiltonian, a Heisenberg interaction term between the
$f$-spins would generate this scale.
As one increases the $f$-spin  fluctuation, one can anticipate that in
the presence of frustration the $f$-spins form a valence bond solid or a
spin-liquid phase. Thereby, they decouple from the conduction electrons
and drop out of the Luttinger  volume count \cite{Beach10, Coleman10} .
}
\end{thebibliography}

\end{document}